\documentclass[pdftex,twocolumn,epjc3]{svjour3}

\RequirePackage[T1]{fontenc}

\smartqed 
\RequirePackage{listings}
\RequirePackage{graphicx}
\graphicspath{{fig/}}
\RequirePackage{tikz}
\RequirePackage{physics}
\RequirePackage{siunitx}
\RequirePackage{flushend}
\RequirePackage[numbers,sort&compress]{natbib}
\RequirePackage[bookmarksopen,bookmarksdepth=3,colorlinks,citecolor=blue,urlcolor=blue,linkcolor=blue]{hyperref}
\usetikzlibrary{arrows, decorations.markings}

\journalname{Eur. Phys. J. C}

\begin{document}

\title{HPGe detector field calculation methods demonstrated with an educational program, GeFiCa}

\author{Jianchen Li \and Jing Liu\thanksref{e2} \and Kyler Kooi}

\thankstext{e2}{e-mail: Jing.Liu@usd.edu}

\institute{Department of Physics, University of South Dakota,\\414 East Clark Street, Vermillion, South Dakota 57069, USA}

\date{Received: date / Accepted: date}

\maketitle

\begin{abstract}
  A review of tools and methods to calculate electrostatic potentials and fields inside high-purity germanium detectors in various configurations is given. The methods are illustrated concretely with a new educational program named GeFiCa - Germanium detector Field Calculator.  Demonstrated in GeFiCa are generic numerical calculations based on the successive over-relaxation method as well as analytic ones whenever simplification is possible due to highly symmetric detector geometries. GeFiCa is written in C++, and provided as an extension to the CERN ROOT libraries widely used in the particle physics community. Calculation codes for individual detectors,  provided as ROOT macros and python scripts, are distributed along with the GeFiCa core library, serving as both examples showing the usage of GeFiCa and starting points for customized calculations. They can be run without compilation in a ROOT interactive session or directly from a Linux shell. The numerical results are saved in a ROOT tree, making full use of the I/O optimization and plotting functionalities in ROOT. The speed and precision of the calculation are comparable to other commonly used packages, which qualifies GeFiCa as a scientific research tool. However, the main focus of GeFiCa is to clearly explain and demonstrate the analytic and numeric methods to solve Poisson's equation, practical coding considerations and visualization methods, with intensive documentation and example macros. It serves as a one-stop resource for people who want to understand the operating mechanism of such a package under the hood.
\end{abstract}

\section{Introduction}
The calculation of electrostatic potentials and fields in a high-purity germanium (HPGe) detector is the initial step in a full pulse-shape simulation~\cite{siggen, adl, salathe15, giovanetti15, pss} process. It is also used to guide the design of novel detector geometries to avoid unreasonably high depletion voltages or hidden undepleted regions, which may occur when the size of a detector is enlarged\cite{david}. The design of read-out electronics can benefit from it as well since the capacitance of a detector, a determination factor of the electronics noise, can be calculated from the energy stored in the electric field in the detector. It is widely used in HPGe detector based neutrinoless double beta ($0\nu\beta\beta$) decay experiments, such as GERDA\cite{gerda} and MJD~\cite{mjd}, dark matter experiments, such as CoGeNT~\cite{cogent}, Texono~\cite{texono} and CDEX~\cite{cdex}, and gamma-ray tracking detectors to study structures of atomic nuclei, such as AGATA~\cite{agata} and GRETA~\cite{greta}, etc.

A complete list is impossible, but commonly used field calculation packages include fieldgen (an essential part of siggen~\cite{siggen, ringberg}) used in GRETITA~\cite{gretina} (an early phase of GRETA) and MJD~\cite{mjd}, SIMION\cite{simion, adl, salathe15, giovanetti15} used in AGATA~\cite{agata} and GERDA~\cite{gerda}, Maxwell\cite{maxwell} used in most experiments, MaGe\cite{mage} used in GERDA~\cite{gerda} and MJD~\cite{mjd}, and FEniCS~\cite{fenics}, a popular open-source computing platform for solving partial differential equations. A new package called \emph{SolidStateDetectors.jl}~\cite{oliver19}, SSD in short hereafter, is under rapid development at the Max-Planck-Institut f\"ur Physik for LEGEND~\cite{legend}, a new $0\nu\beta\beta$ experiment as a combined effort of GERDA and MJD.

SIMION is a commercial software package primarily used to simulate the transportation of charged particles in static or low-frequency RF fields. According to its documentation~\cite{simion}, it uses the finite-element method to calculate 2D and 3D fields with up to almost 20 billion grid points, given enough RAM. Its power in static field calculation is overkill for HPGe detectors while lacking some important features that are required for HPGe detector applications, such as the calculation of depletion voltage, region and detector capacitance, etc. This is understandable given that the main application of SIMION is not HPGe detector field calculation.

ANSYS Maxwell~\cite{maxwell} is a more popular electromagnetic field simulation software compared to SIMION. It is for the design and analysis of electric motors, actuators, sensors, transformers and other electromagnetic and electromechanical devices. It uses automatic adaptive meshing techniques to achieve user-specified accuracy without detailed instruction from a user. As its main application is not HPGe detector field calculation, it has the same advantages and disadvantages as SIMION, but is more expensive than SIMION.

A common pitfall of all general-purpose commercial software is that one has to pay for extra features that are not needed in the HPGe field calculation, while still missing out some basic features that are needed.

FEniCS~\cite{fenics}, on the other hand, is a free-to-use, open-source program developed by a global community of scientists and software developers, and is just as sophisticated as SIMION and Maxwell.  Using efficient finite-element codes, its main purpose is to solve partial differential equations, including Poisson's equation, which is needed in HPGe field calculations. As versatile as it is, FEniCS demands effort to adapt it to a specific application, such as calculating fields in HPGe detectors. From this point of view, FEniCS has the same drawback as commercial packages, that is, it is overkill for HPGe field calculation, but lacks basic features that are specific for HPGe application. Nonetheless, there is ongoing effort within the MJD collaboration to adapt it for HPGe detectors.

On the contrary, MaGe~\cite{mage} and siggen~\cite{siggen, ringberg} are dedicated software for HPGe signal formation simulation. They are not as versatile and sophisticated as the previously mentioned packages, but are sufficient for the HPGe application. Initially, MaGe was jointly developed by the Majorana~\cite{mjd} and GERDA collaborations mainly as a GEANT4~\cite{g41,g42,g43} based Monte Carlo simulation package. It was extended later on to include a full pulse-shape simulation chain using GEANT4 simulation results as input~\cite{jing, daniel, pss}. It can be used for the simulation of both segmented~\cite{abt07} and point-contact~\cite{luke89} detectors. The major drawback of MaGe is that it is only available for the GERDA or Majorana collaborators.

Siggen~\cite{siggen, ringberg} is mainly developed by David Radford for MJD pulse-shape simulation. It is open-source and free to use. A stand-alone portion of siggen, called fieldgen, is dedicated to the calculation of fields and potentials of point-contact detectors in two dimensional cylindrical coordinates. It cannot be used for segmented detectors. The program is written in c, but the configuration file is in plain ASCII with straightforward syntax for a user to easily specify detailed dimensions of a detector, such as the size of small electronic contact, or the width of a groove to reduce surface leakage current. Fieldgen can also be used to calculate the capacitance of a detector, the full depletion voltage, and the depletion region in case that a detector is not fully depleted. Those functions are not available in the packages mentioned previously. Fieldgen utilizes the successive over-relaxation method (SOR) to first calculate the potential in a coarse grid with a typical distance of 1~mm between two grid points. The result of this coarse calculation is then used as the input of a more precise calculation in a finer grid with a typical distance of 0.1~mm between two grid points. Using this simple approach in place of automatic adaptive meshing techniques used in some of the other packages makes fieldgen both fast and accurate enough for its dedicated application.

SSD~\cite{oliver19} is mainly developed by the GeDet group at the Max-Planck-Institut f\"ur Physik for LEGEND~\cite{legend}. It is capable of not only the calculation of electric fields but also the simulation of electronic signals. In its field calculation part, it contains functions to deal with common detector configurations, such as point-contact and segmented ones. It features adaptive grid sizes, which improves both the calculation speed and accuracy. It is written in \emph{Julia}~\cite{julia}, a relatively new, high-level, general-purpose programming language designed to address the needs of high-performance numerical analysis. It is possible to enable multi-threading in SSD, which takes the full advantage of modern computer hardware. Compared to MaGe and fieldgen, SSD has an attractive feature to calculate the field outside of a detector taking into account the influence of the detector holding structure nearby.

Overall, fieldgen seems to be the maturest at this moment for users interested in HPGe field calculation as long as their detector geometry is similar to that of point-contact ones~\footnote{It is not as limiting as it sounds, because the bore hole of a common coaxial detector can be regarded as a very large point-contact, and a planar detector can be regarded as having a large flat point-contact. The main limitation is on segmented or stripped contacts.}. However, the lack of detailed documentation makes it hard for a developer to modify the code of fieldgen for other geometries or to add new features.

This is one of the reasons why many research groups write their own code for HPGe detector field calculation instead of using the mentioned major players. An obvious advantage of home brewed code is that it is well understood and easy to tune if needed. The second advantage is that writing their own code instead of using existing ones deepens the understanding of junior researchers on HPGe detector working principles and numerical calculation techniques. Drawbacks of this approach include the limited functionality, the lack of verification and the waste of time in reinventing the wheel.

GeFiCa is aimed at clear explanation and demonstration of the analytic and numeric methods to solve Poisson's equation, practical coding considerations and visualization methods. It does so by providing intensive documentation and example macros, and serves as a one-stop resource for people who want to understand the operating mechanism of such a package under the hood. None of the tools mentioned above fits all applications. Home brewed codes built on top of some existing tools may be the best choice for education and specific applications, as long as the drawbacks mentioned previously can be effectively overcome through the demonstration provided in GeFiCa.

\section{Space charges}
\label{s:sc}
\begin{figure}[htbp] \centering
  \includegraphics[width=\linewidth]{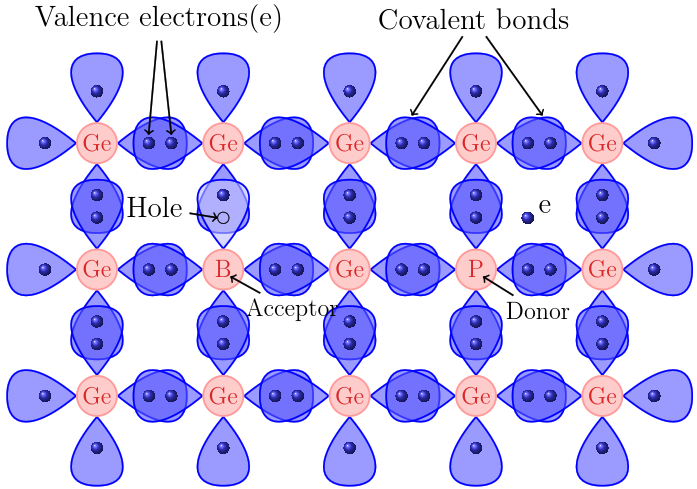}
  \caption{Conceptual sketch of covalent bonds between Ge and impurity atoms (P and B, as examples).}
  \label{f:bond}
\end{figure}

HPGe crystals come in two types. As shown in Fig.~\ref{f:bond}, if the trace impurity atoms in a crystal provide free-moving electrons (phosphorus, for example), the crystal is of $n$-type, and if the atoms provide free-moving holes (boron, for example), the crystal is of $p$-type. In both literature and popular science articles, these free-moving charge carriers are often preceded with adjectives like ``extra'' or ``excess'', which may lead to a false impression that an $n$-type crystal has ``extra'' electrons donated by donor impurity atoms and is hence negatively charged, or that a $p$-type crystal has ``extra'' holes (vacancies in covalent bonds) due to acceptor impurity atoms and is positively charged. These free-moving charges can be regarded as ``extra'' since they are not used in forming covalent bonds between atoms, which is the fundamental reason why they are free. But they are not ``extra'' charges that break the balance of the numbers of protons and electrons in a crystal.  Actually, no matter which type it is, a crystal is electrically neutral as a whole because the number of protons are the same as the number of electrons in both impurity and Ge atoms. As trivial as it sounds, this fact is worthy of emphasizing, especially for one to understand the sign of space charges to be mentioned in the following paragraph.

When the bias voltage applied to a crystal is high enough, all free-moving charge carriers can be swept out of the bulk of the crystal. The crystal is said to be depleted of free charge carriers. In an $n$-type crystal, it is the free-moving electrons that are swept out. Consequently, the trace impurity atoms are positively ionized. Since the ions are fixed in their locations in the crystal, they cannot be swept out by the external electric field, and are hence called ``space charges''. In a $p$-type crystal, however, the space charges are negative, since it is the free-moving holes that are swept out.  It is quite counter intuitive for one to realize the fact that a depleted $n$-type crystal is actually positively charged and a $p$-type negatively charged. Space charges create an electric field in addition to the one that is created by the bias voltage. The total electric field inside a depleted crystal is a linear combination of these two.

The space charge density distribution, normally denoted as $\rho$, can be quite complicated due to the nature of HPGe single crystal growth process~\cite{hansen82, wang15}. It is normally characterized in the following way. First, a few wafers are cut from various axial positions in a HPGe single-crystal boule pulled using the Czochralski method, typically, one from the shoulder and one from the tail of the boule. Second, small samples are cut from individual wafers along their radius.  Net impurity concentrations of these samples, $N_A-N_D$, are then measured using Hall-effect, where $N_A$ is the acceptor concentration, $N_D$ donor concentration, with a unit of cm$^{-3}$. Since there is a relationship between $\rho$ and  $N_A-N_D$ as explained in the previous paragraph:
\begin{equation}
  \rho=-(N_A-N_D)e,
\end{equation}
where $e=1.6\times10^{-19}$~C is the elementary charge, both the vertical (axial) and radial distributions of $\rho$ can be investigated this way.

\begin{figure}[htbp] \centering
  \includegraphics[width=\linewidth]{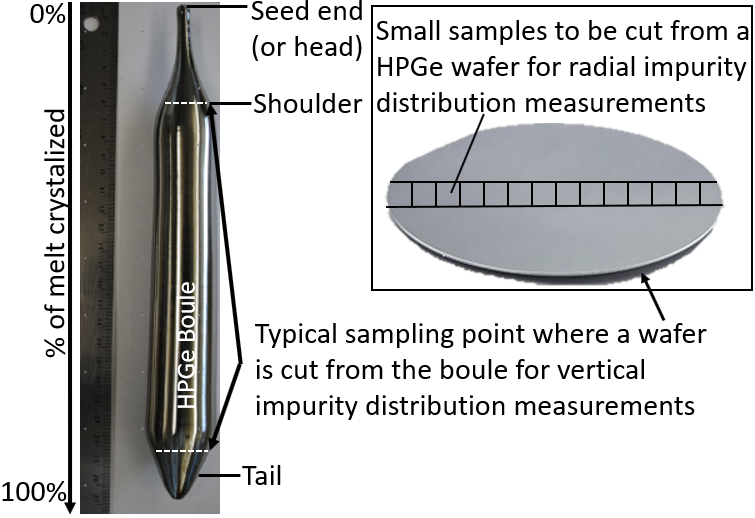}
  \caption{A HPGe single-crystal boule pulled using the Czochralski method, and a HPGe wafer cut from the boule for impurity measurements.}
  \label{f:xstl}
\end{figure}

\begin{figure}[htbp] \centering
  \includegraphics[width=\linewidth]{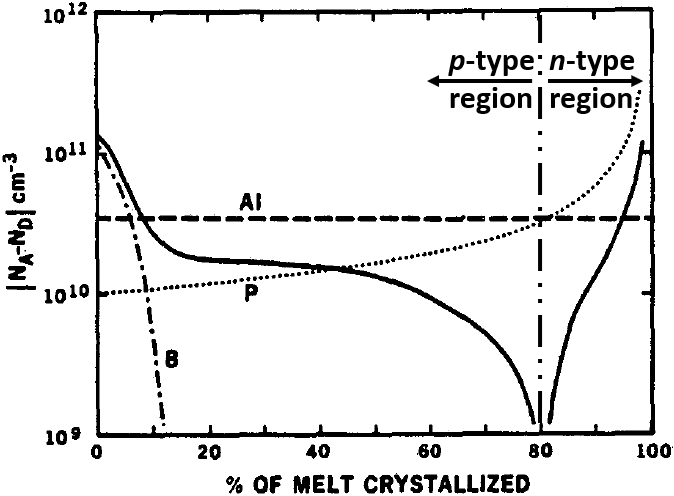}
  \caption{A typical vertical net impurity concentration profile of a HPGe single-crystal boule, taken from \cite{hansen82}.}
  \label{f:prof}
\end{figure}

A typical vertical net impurity concentration profile of a HPGe single-crystal boule is shown in Fig.~\ref{f:prof} taken from \cite{hansen82} with a vertical double-dotted dashed line added to clearly indicate the $p$-type and $n$-type regions. The dashed line indicates the contribution from a typical $p$-type impurity element, Al. The dotted dashed curve indicates the contribution from another typical $p$-type impurity element, B. The dotted curve shows the contribution from a typical $n$-type impurity element, P. The solid curve broken around 80\% of the boule is the overall net impurity concentration. The crystal is of $p$-type from 0 to 80\% of its length, and changes to $n$-type after that. The curve is approximately flat from 20\% to 40\%, which is a typical portion of the boule to be harvested for detector fabrication.

\begin{figure}[htbp] \centering
  \includegraphics[width=\linewidth]{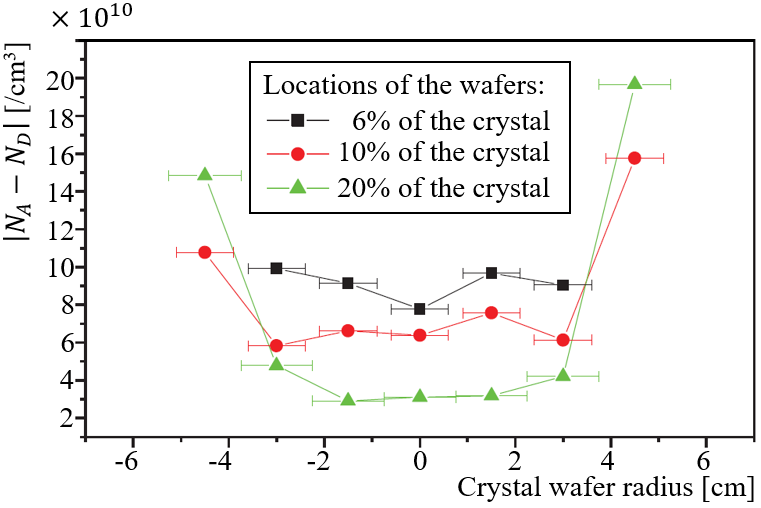}
  \caption{A typical radial net impurity concentration profile of a HPGe single-crystal boule, taken from \cite{wang15}.}
  \label{f:rpro}
\end{figure}

A typical radial net impurity concentration profile of a HPGe single crystal is shown in Fig.~\ref{f:rpro} taken from \cite{wang15}. It is basically flat from 0 to a certain radius, but increases dramatically close to the skin of the crystal. Sometimes, a crystal may even change its type from its center to its outer radius, as mentioned in \cite{hansen82}. The skin of a boule may be removed so that the central part used for detector fabrication has a relatively constant impurity distribution.

Given those experimental evidences, the space charge density in general has to be expressed as a function of location, \textit{i.e.}, $\rho(\vec{x})$, where $\vec{x}$ is a vector indicating the location of interest. Since the measurement of impurity is destructive for the raw material, the real impurity distribution in a crystal used for detector fabrication is usually unknown. Normally, only the average impurities close to the top and bottom of the cut portion of the crystal are known. The impurity distribution in between is regarded as a constant or approximated by a first-order polynomial determined by the average top and bottom impurities. If the right portion of a crystal (20--40\% of the black line in Fig.~\ref{f:prof}, for example) is harvested for detector fabrication, this is normally an acceptable approximation. However, one has to keep in mind that our knowledge of the real impurity distribution is incomplete, and our approximation may have sizable uncertainties.

\section{Poisson's Equation}
The existence of space charges complicates the calculation of the electrostatic potential in a HPGe detector. Without space charges, the potential can be calculated by solving Laplace's equation,
\begin{equation}
  \label{e:lap}
  \nabla^2 V(\vec{x}) = 0,
\end{equation}
where $V(\vec{x})$ is the potential to be determined. With space charges, however, the potential must be calculated by solving Poisson's equation, which takes into account the space charge distribution in the bulk of a detector:
\begin{equation}
  \label{e:poe}
  \nabla^2 V(\vec{x}) = -\frac{\rho(\vec{x})}{\epsilon},
\end{equation}
where $\epsilon=\epsilon_0\epsilon_r$ with $\epsilon_0 \approx 8.854 \times 10^{-12}$~F/m being the permittivity in vacuum, and $\epsilon_r \approx 16.0$ being the relative permittivity (or dielectric constant) of Ge.

Both differential equations have an infinite amount of solutions characterized by a few undetermined constants. These constants can be fixed by boundary conditions, which refers to the voltage values on detector electrodes. The relationship between these two equations can be understood better when we consider two different boundary condition setups: first, potentials of electrodes of a detector are set based on the bias voltage applied to the detector, second, they are all set to zero. If Laplace's equation is solved with the first setup, its solution is a potential field caused by the bias only. If Poisson's equation is solved with the second setup, its solution is simply the potential caused by the space charges only. The potential in a detector is a linear combination of these two solutions. We can also solve Poisson's equation with the first set of boundary conditions, which directly results in the combined potential.

In addition to the potential, we are also interested in the electric field distribution in a detector. The electric field vector $\vec{E}$ can be then determined with the equation
\begin{equation} \label{e:edv}
  \vec{E}=-\nabla V.
\end{equation}

These equations are rather abstract. A concrete expression can be obtained in a specific coordinate system. For example, in Cartesian coordinates, Poisson's equation reads,
\begin{equation}
  \label{e:pxyz}
  \frac{\partial^2 V}{\partial x^2}+\frac{\partial^2 V}{\partial y^2}+\frac{\partial^2 V}{\partial z^2}=-\frac{\rho(x,y,z)}{\epsilon},
\end{equation}
where, $x, y, z$ are the three Cartesian coordinates. The expression of Poisson's equation in spherical and cylindrical coordinates are listed in \ref{a:pesc}.

\section{Analytic Solutions}
\subsection{Planar Detectors}

\begin{figure}[htbp] \centering
  \includegraphics[width=0.5\linewidth]{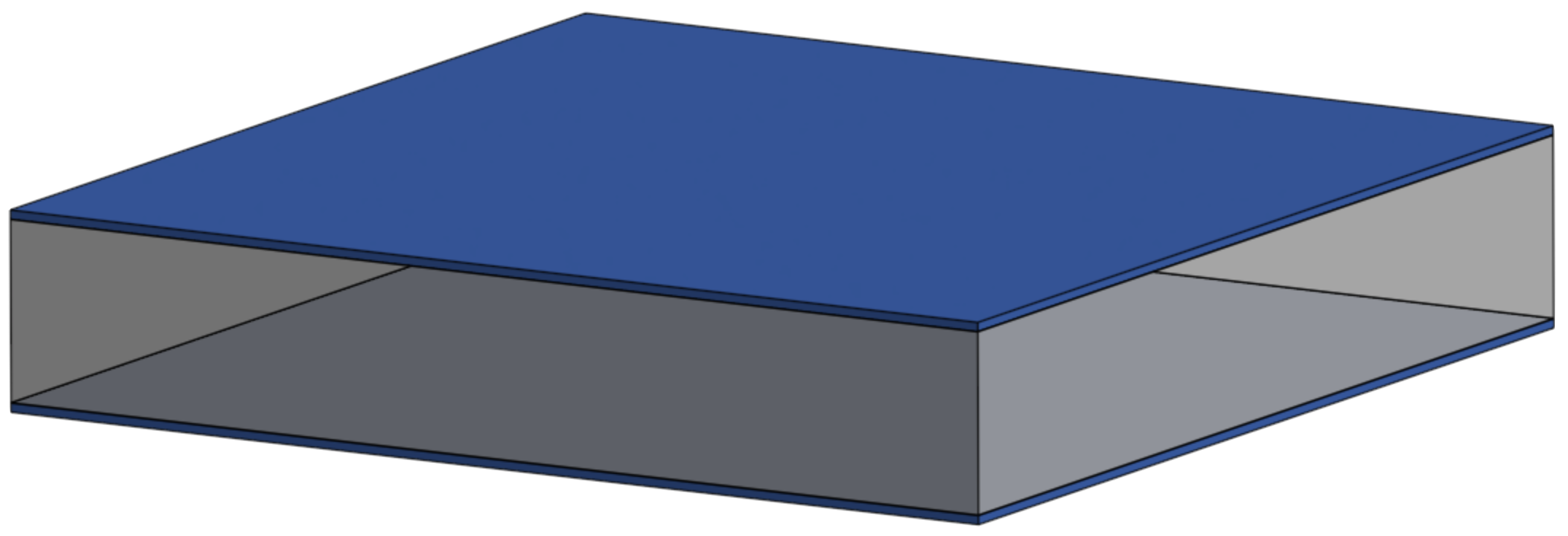}
  \caption{3D model of a planar detector with electrodes indicated with blue.}
  \label{f:plan}
\end{figure}

As mentioned in section~\ref{s:sc}, in general, the space charge density $\rho$ is a function of $\vec{x}$, and there is no analytic solution for three dimensional Poisson's equation with a complicated $\rho(\vec{x})$. However, in certain highly symmetric detector configurations, Poisson's equation can be significantly simplified and its analytic solution can be obtained. For example, at the center of a large but thin planar HPGe detector, the electric potential can be regarded as only varying along the thickness of the detector, $x$, and $\rho$ can be regarded as a constant. Eq.~\ref{e:pxyz} can then be simplified to
\begin{equation}
  \label{e:px}
  \dv[2]{V}{x}=-\frac{\rho}{\epsilon}.
\end{equation}
Its analytic solution reads,
\begin{equation}\label{e:asp}
  V=-\frac{\rho}{2\epsilon}x^2+C_2x+C_1,
\end{equation}
where $C_1$ and $C_2$ are constants which can be determined using two boundary conditions, \textit{i.e.}, the voltages of two planar detector electrodes. The electric field reads,
\begin{equation}\label{e:aspe}
  E=-\dv{V}{x}=\frac{\rho}{\epsilon}x-C_2,
\end{equation}

\begin{figure}[htbp]
  \includegraphics[width=\linewidth]{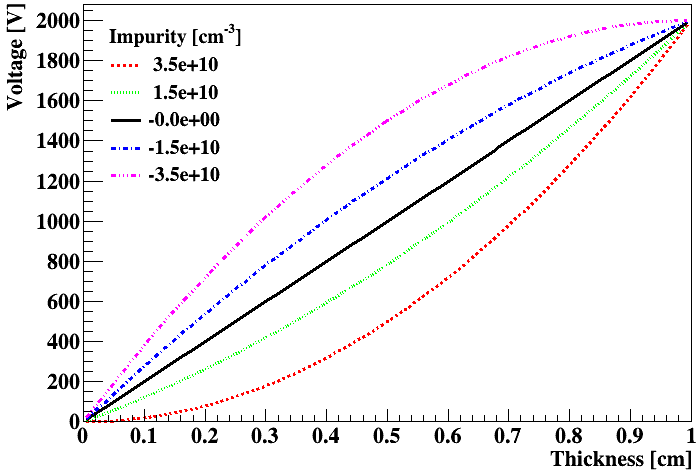}
  \caption{Voltage versus thickness of a planar detector.}
  \label{f:vx}
\end{figure}

\begin{figure}[htbp] \centering
  \includegraphics[width=\linewidth]{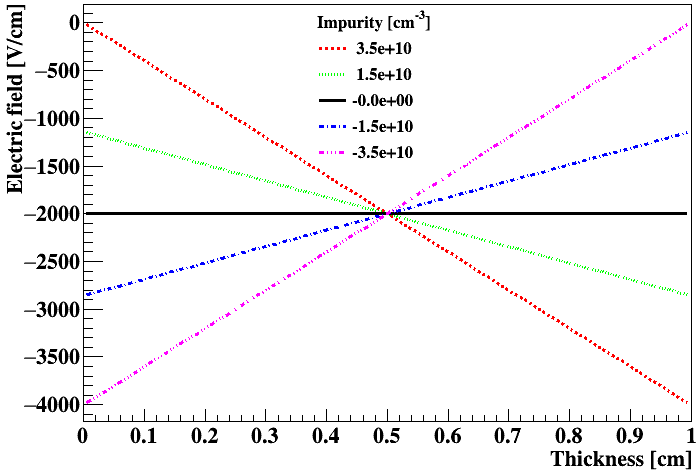}
  \caption{Electric field versus thickness of a planar detector.}
  \label{f:ex}
\end{figure}

Fig.~\ref{f:vx} shows the voltage versus the thickness of a planar detector, assuming a thickness of 1~cm and a voltage of 2,000~V applied to its top electrode. The net impurity concentration corresponding to each curve in the figure is listed in the legend. When $\rho=0$, Eq.~\ref{e:asp} becomes $V=C_2 x + C_1$, which is simply a straight line between [0,0] and [1~cm, 2,000~V]. The higher an impurity concentration, the more a curve is bent up or down depending on the type of the impurity. Since the slope of curves in Fig.~\ref{f:vx} is proportional to the magnitude of the electric field, as shown in Eq.~\ref{e:aspe}, the bending of the curves shows how space charges modify the overall electric field in a detector. This is demonstrated explicitly in Fig.~\ref{f:ex}, where the $y$-axis changes to electrical field in the unit of V/cm. A small change of the impurity concentration may result in large deviation of the overall field from the constant external field. When the impurity concentration is high enough, the electric field close to the electrodes of the detector can be as low as zero. Such low field regions are where severe charge trapping may happen, which deteriorates the energy resolution of the detector, hence are not desirable. Obvious solutions of such a problem include applying a voltage significantly higher than the depletion voltage, reducing the thickness of the detector, or growing purer crystals. The first solution is dangerous, the second is undesirable and the last is difficult. A less obvious alternative is to switch to a different geometric configuration of the detector.

\subsection{Coaxial Detectors}

\begin{figure}[htbp] \centering
  \includegraphics[width=0.3\linewidth]{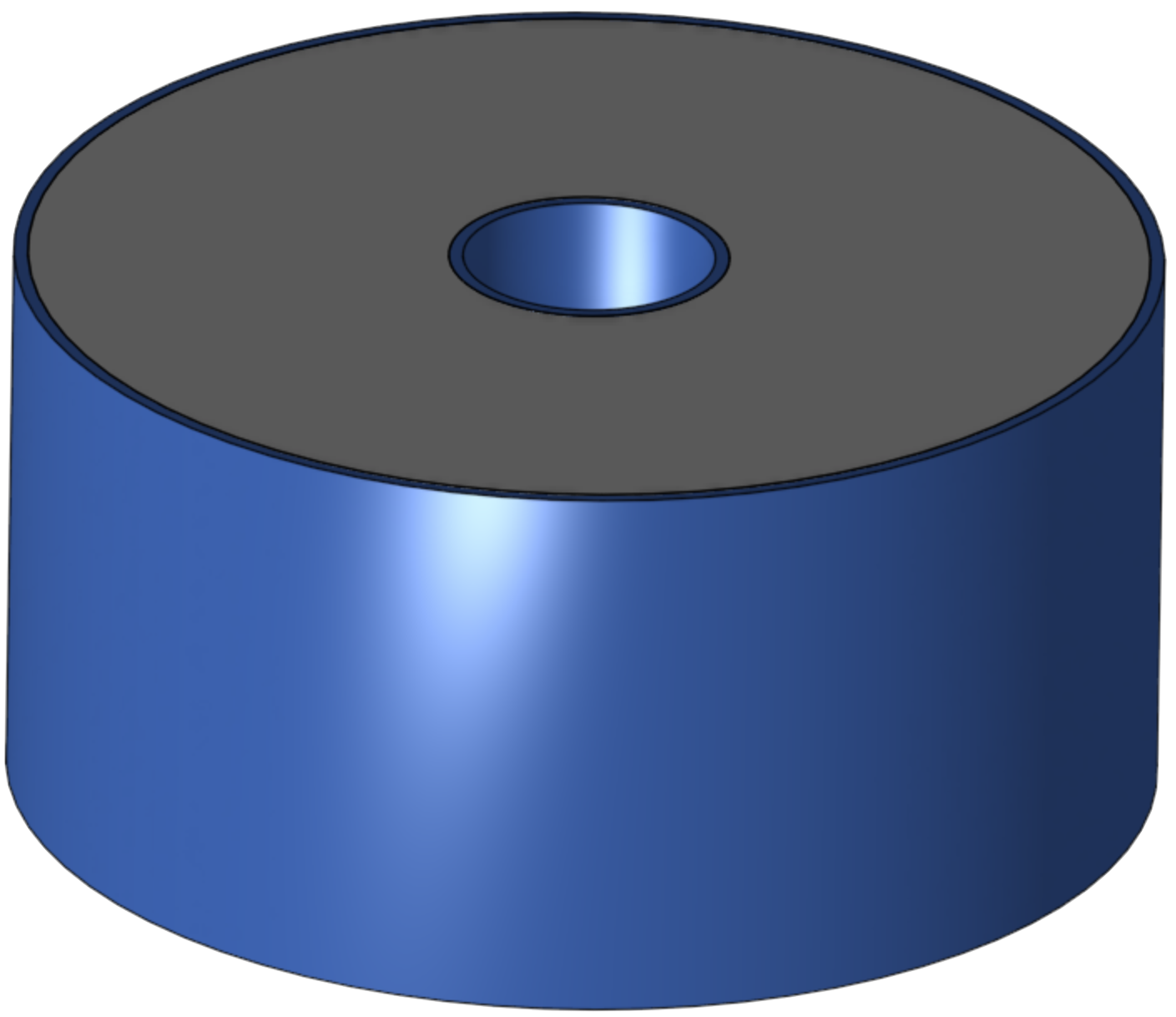}
  \caption{3D model of true coaxial detector with electrodes indicated with blue.}
  \label{f:coax}
\end{figure}

How the geometry of a detector can help solve this problem can be clearly demonstrated using the analytic solution of Poisson's equation in cylindrical coordinates.  The electric potential far away from its two end surfaces of a true-coaxial HPGe detector can be regarded as varying only with $r$. If one further assumes that the space charge density $\rho$ is a constant, Poisson's equation in cylindrical coordinates can be simplified to
\begin{equation}
  \frac{1}{r}\dv{}{r}\left(r\dv{V}{r}\right)
  =-\frac{\rho}{\epsilon}.
\end{equation}
Its analytic solution reads,
\begin{equation}
  V=-\frac{\rho r^2}{4\epsilon}+C_1log(r)+C_2,
\end{equation}
where $C_1$ and $C_2$ are constants, which can be determined using boundary conditions, that is, the locations and voltages of the two electrodes of a detector. The electric field is then
\begin{equation}
  \vec{E}=-\nabla V = \frac{\rho r}{2\epsilon} - \frac{C_1}{r}.
\end{equation}

\begin{figure}[htbp] \centering
  \includegraphics[width=\linewidth]{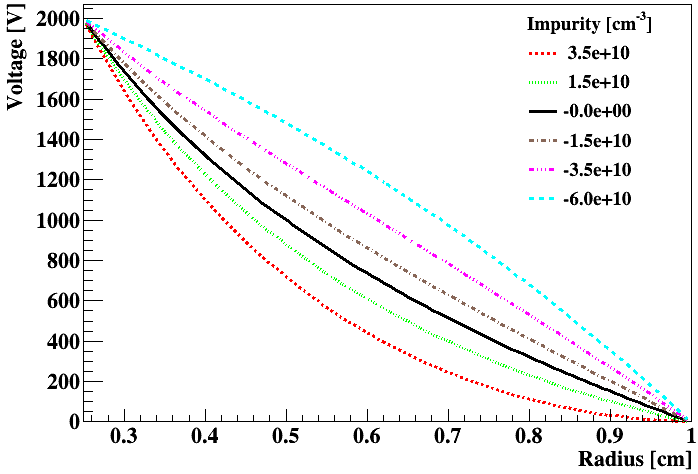}
  \caption{Voltage versus radius of a true coaxial detector.}
  \label{f:vr}
\end{figure}

\begin{figure}[htbp] \centering
  \includegraphics[width=\linewidth]{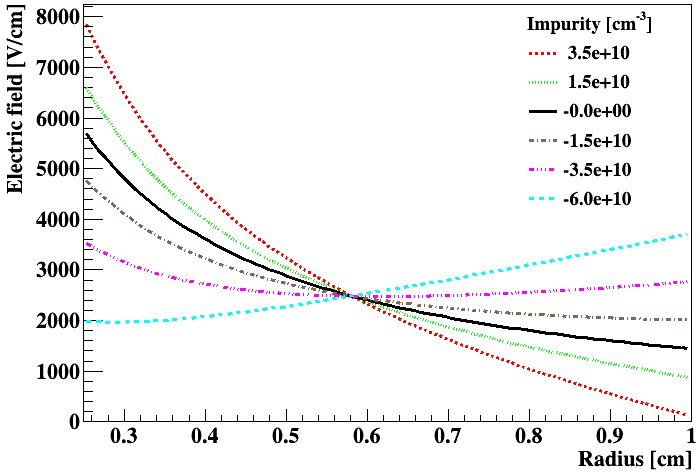}
  \caption{Electric field versus radius of a true coaxial detector.}
  \label{f:er}
\end{figure}

Fig.~\ref{f:vr} shows the voltage versus the radius of a true-coaxial detector with an inner radius of 0.25~cm, an outer radius of 1~cm and a voltage of 2,000~V applied to its inner electrode. The net impurity concentration corresponding to each curve in the figure is listed in the legend. Compared to Fig.~\ref{f:vx} for a planar detector, the curve corresponding to the zero impurity is not a straight line anymore. Instead, it bends downward, reflecting the fact that the electrical field close to the inner radius is stronger, which can be seen in Fig.~\ref{f:er} as well. Now, $p$-type impurities bend the curves further down, while $n$-type impurities bend them upwards, effectively flatten the electrical field distributions along the radius, as shown explicitly in Fig.~\ref{f:er}. Given the right impurity concentration, the electrical field in a true-coaxial detector can be optimized to avoid charge trapping. To demonstrate this point more clearly, a curve corresponding to a high impurity concentration of $-6\times10^{10}$/cm$^3$ is added to Fig.~\ref{f:vr} and \ref{f:er}, which is not in Fig.~\ref{f:vx} and \ref{f:ex}. The electric field distribution corresponding to this concentration is in between $\sim$2,000~V/cm and 4,000~V/cm, well above zero in the entire volume of the detector. One has to avoid falling into a false impression that coaxial detectors prefer $n$-type crystals to $p$-type ones. In reality, this preference can be easily flipped by flipping the bias polarity. It is better to say that the type of the crystal prefers a certain bias polarity. As a conclusion, coaxial detectors are much more tolerable for high impurity concentrations than planar ones.

\subsection{Hemispherical Detectors}

\begin{figure}[htbp] \centering
  \includegraphics[width=0.4\linewidth]{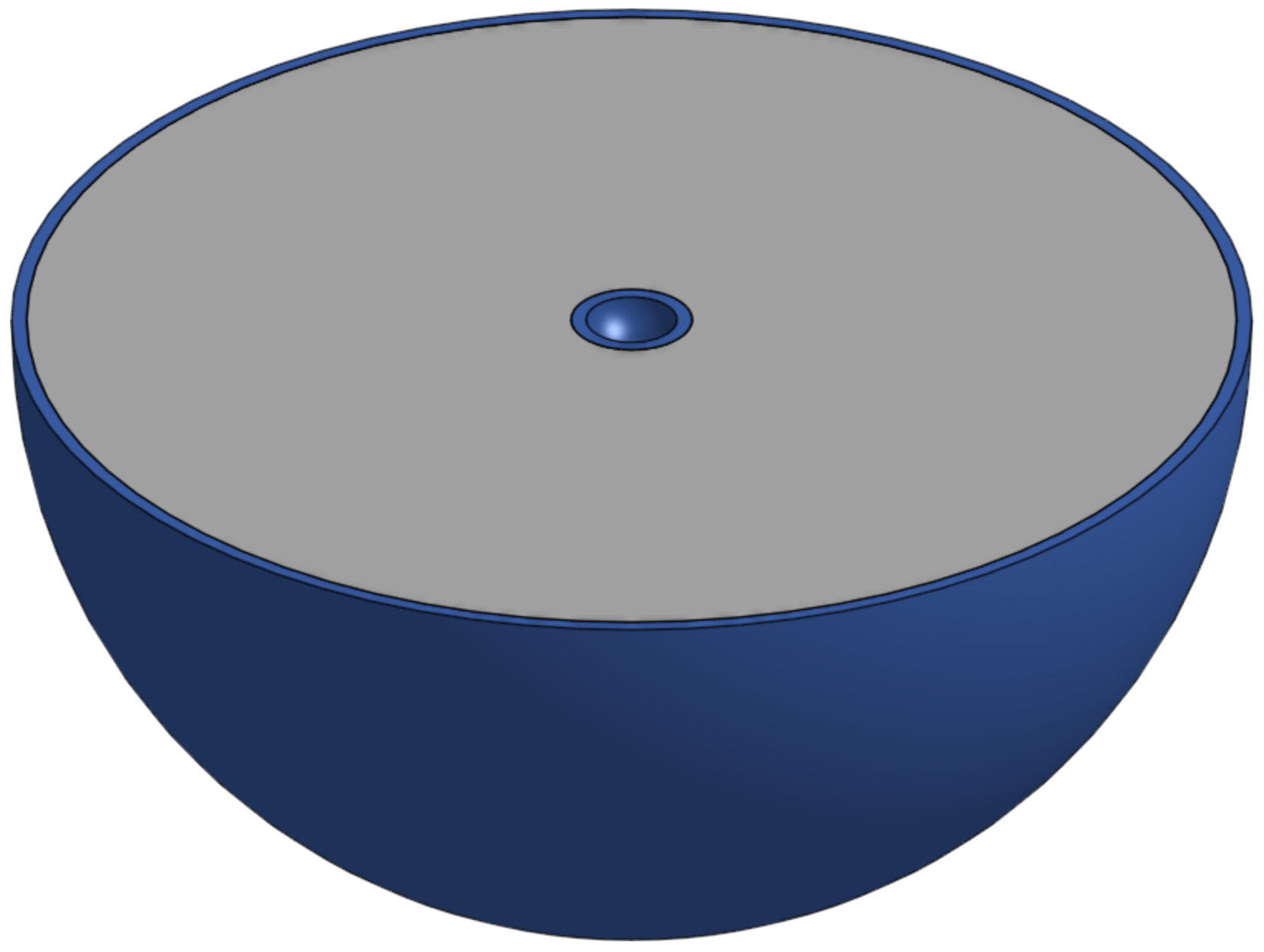}
  \caption{3D model of a hemispherical detector with electrodes indicated with blue.}
  \label{f:hemi}
\end{figure}

In spherical coordinates with polar and azimuthal symmetries, the $\theta$ and $\phi$ terms can be dropped and the Poisson's equation can be simplify to
\begin{equation}
  \frac{1}{r^2}\dv{}{r}\left(r^2\dv{V}{r}\right)
  =-\frac{\rho}{\epsilon}.
\end{equation}
Assuming constant $\rho$, its analytic solution reads,
\begin{equation}
  V=-\frac{\rho r^2}{6\epsilon}+C_1\frac{1}{r}+C_2,
\end{equation}
where $C_1$ and $C_2$ are constants that can be determined by boundary conditions. The electrical field
\begin{equation}
  \vec{E}=-\nabla V = \frac{\rho r}{3\epsilon}+C_1\frac{1}{r^2}.
\end{equation}
A detector with such a configuration is more impurity tolerable than a coaxial one. However, it is not easy to bias the inner radius of a full sphere. The electrical field of a hemispherical detector can be approximated with the same solution, and it is possible to apply voltage to its inner radius. However, it is a significant challenge to machine a cylindrical single-crystal boule into such a shape. For this reason, no hemispheric HPGe detectors have been made so far, and most HPGe detectors take the cylindrical shape for convenience.

If the inner radius of an imaginary hemispheric detector is small enough, say, about 1~mm, it can help illustrate some important properties of a point-contact detector, which can be imagined as a traditional coaxial detector with its central contact shrunk to a point. For example, certain amount of impurity is necessary to shape the electrical field in the detector so that it is not too strong close to the point-contact and not too weak far away from it. This is why the first point-contact detector is called a ``shaped-field'' one~\cite{luke89}.

\subsection{Depletion Voltage}
\label{s:adv}
Given fixed dimensions and impurity concentration of a crystal, we'd like to find out the voltage at which the crystal can be fully depleted, or the depletion voltage, $V_\text{d}$. The method to solve this problem can be demonstrated using the analytic solution, Eq.~\ref{e:asp}, of the one-dimensional Poisson's equation in Cartesian coordinates. The strategy can be applied to multi-dimensional configurations with minor modifications.

\begin{figure}[htbp] \centering
  \includegraphics[width=\linewidth]{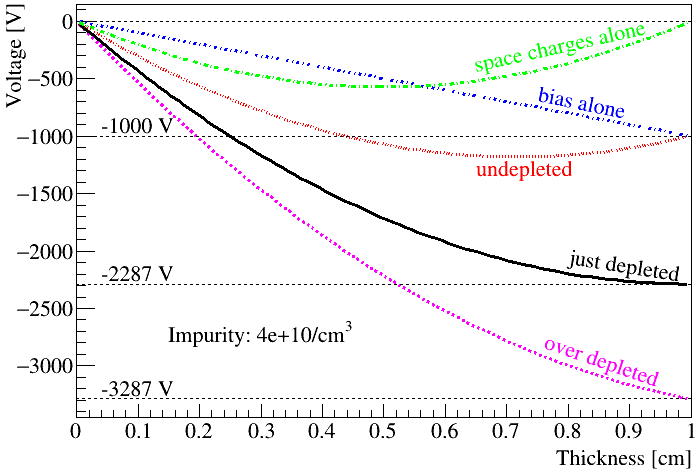}
  \caption{Over depleted, just depleted, and undepleted voltage distributions in case of an ideal planar detector.}
  \label{f:und}
\end{figure}

To keep our discussion as concrete as possible, let us assume an ideal planar detector with a thickness of $d=1$~cm and a homogeneous impurity concentration of $4\times10^{10}$/cm$^3$ ($p$-type) in its entire volume.  Let's further assume that its bottom electrode is at $x=0$ and grounded, i.e.
\begin{equation}\label{e:b1}
  V(x=0)=0
\end{equation}
Applying this boundary condition to Eq.~\ref{e:asp}, we have $C_1=0$. If no bias is applied at the top electrode, that is, $V(x=d)=0$, we can further get $C_2=\rho d/(2\epsilon)$, where $\rho=-4\times10^{10}$/cm$^3 e$ is the space charge density. Eq.~\ref{e:asp} can then be rearranged as,
\begin{equation}\label{e:para}
  V(x) = -\frac{\rho}{2\epsilon}(x-\frac{d}{2})^2+\frac{\rho d^2}{8\epsilon},
\end{equation}
which is the green parabola shown in Fig.~\ref{f:und}. However, this is not physically correct, since the whole crystal should be at $V=0$ without any bias. The problem comes from our taken-as-granted assumption that $\rho=-4\times10^{10}$/cm$^3 e$, which is only true in depleted region. In undepleted region, there should not be any space charge, that is, $\rho=0$, which indeed guarantees $V(x)=0$ in Eq.~\ref{e:para}.

Instead of regarding it as a mistake, there is a better way to interpret Eq.~\ref{e:para}, that is, it is the contribution to the voltage from the ``space charge alone'', when the detector is fully depleted. Its value hence is not dependent on the bias voltage after fully depletion. The overall voltage should be the sum of this contribution and the voltage due to the external bias.

The external bias voltage distribution without any contribution from space charges is simply $V(x)=C_2 x$ (Eq.~\ref{e:asp} with $\rho=0, C_1=0$), that is, a straight line, as the one labelled ``bias alone'' in Fig.~\ref{f:und}.

At a bias voltage of $-1000$~V, the sum of ``space charge alone'' and ``bias alone'' contributions gives the red curve in Fig.~\ref{f:und} that is below the $-1000$~V line in a wide region. This region can be regarded as a potential well where positive free charge carriers are trapped. In another word, the $-1000$~V bias is not enough to sweep all free charges out of the crystal. The curve is hence labelled ``undepleted''.

What we are interested in here is to identify differences between an undepleted case and a depleted one. In this concrete example, an ``undepleted'' curve has $|V(x)|>|V_\text{d}|$ at some $x$. If we change the crystal from $p$-type to $n$ and keep other configurations unchanged, the ``space charge alone'' curve would bend downward. More general criteria hence would be, an ``undepleted'' curve has $V(x)$ out of the range defined by boundary voltages, or $\dd{V}/\dd{x}$ changing its sign, at some $x$.

Assuming a certain bias voltage and a constant $\rho$ over the whole volume, if the final answer is ``undepleted'' according to the criteria identified previously, we have to start over again assuming a higher bias. Obviously, the detector will certainly be depleted given an extremely high bias. However, in reality, it is hard to deliver a very high voltage without micro (or even major) discharges along the high voltage cable. Normally, the operation voltage is $\sim 1000$~V over the depletion voltage.

The difference between an over depleted curve and a just depleted one, in this concrete example, is that
\begin{equation}\label{e:b2}
E(x=d)=-\dd{V(x=d)}/\dd{x}=0
\end{equation}
for the latter, but $E(x=d)>0$ for the former case.

In general, we have to do a search in between 0 and a large bias voltage for the ``just depleted'' case, where the electric field $E$ on one of the boundaries is exactly zero. The calculation for this analytic example is very fast. Special treatment has to be taken in multi-dimensional numerical calculations to avoid expensive computations (See Sec.~\ref{s:dep}).

\subsection{Impurity Requirement}
Given technical difficulties in delivering high voltages in a cryogenic environment, a low depletion voltage is always preferred. It is a common practice to figure out the maximal net impurity concentration a crystal with certain dimensions must have to be depleted at or under a given voltage. In our previous example, the depletion requirement is $E(d)=0$, which allows us to calculate $C_2$:
\begin{equation*}
  E(x=d)=\dv{V(d)}{x}=-\frac{\rho}{\epsilon}d+C_2=0 \Rightarrow C_2=\frac{\rho}{\epsilon}d.
\end{equation*}
Insert the calculated $C_2$ and $C_1=0$ back to Eq. ~\ref{e:asp}, we get the maximal allowed space charge concentration $\rho=2V\epsilon/d^2$, where $V$ is the given voltage.

Analytic solutions are not available for more complicated detector configurations. In that case, we need to make guesses on the impurity concentration or even profile, search for corresponding depletion voltages based on the method described in Sec.~\ref{s:dep}, and see if they go below the required voltage.

\section{Numerical Calculation}
Even though many detector design concepts can be demonstrated with analytic solutions of highly symmetric detector configurations, numerical calculations are necessary for more advanced configurations that cannot be simplified to lower dimensional problems.

The first step of numeric calculation is to establish a grid within the detector volume, which consists of many tightly spaced points, some right on boundaries, others inside.
The field values of a grid point can be determined by those of its immediate neighboring points. Their relations are dictated by Poisson's Equation in its numeric forms. Starting with the known values of the points on boundaries, the value of each point can be uniquely determined.

\begin{figure}[htbp] \centering
  \includegraphics[width=0.5\linewidth]{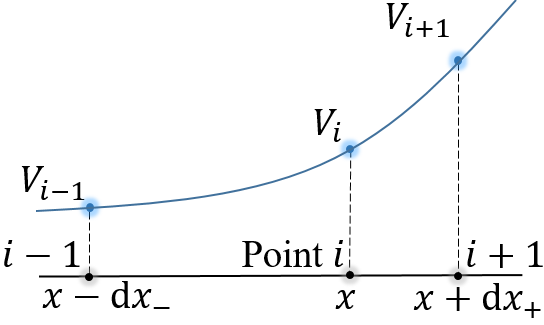}
  \caption{A section of a 1D grid around a point at $x$.}
  \label{f:grid}
\end{figure}

Configuring a grid that ensures an efficient and accurate calculation is an art by itself. For the sake of clarity in our discussion without losing generality, let's at first consider a section of a one dimensional (1D) grid around a point at $x$, as shown in Fig.~\ref{f:grid}.

Numerically, the second order derivative on the left side of Eq.~\ref{e:px} can be expressed as
\begin{equation}\label{e:ndv}
  \dv[2]{V}{x}=\frac{(V_{i+1}-V_i)/\dd x_+ - (V_i- V_{i-1})/\dd x_-}{(\dd x_++\dd x_-)/2},
\end{equation}
where $\dd x_\pm$ are the distances from the point at $x$ to the previous and the next points as shown in Fig.~\ref{f:grid}. It is possible to involve more points in the calculation, such as the previous previous or next next points, but the basic idea is the same.

There are different ways to rearrange Eq.~\ref{e:px} based on Eq.~\ref{e:ndv}, which lead to different methods to solve the problem. The two most common ones are the conjugate gradient method and the successive over-relaxation method.

\subsection{Conjugate Gradient Method}
The conjugate gradient method starts by moving all known terms, such as the boundary voltages and terms containing $\rho(x)$, etc., to the right side of Eq.~\ref{e:px}. Assuming $n$ points in our 1D grid shown in Fig.~\ref{f:grid}, and $V(x)=V_i$, $\rho(x)=\rho_i$ are the values at the i$^{th}$ point, Eq.~\ref{e:px} becomes,
\begin{equation*}
  0\cdot V_1 + \cdots + C_{i-1}V_{i-1} + C_iV_i + C_{i+1}V_{i+1} + \cdots 0 \cdot V_{n-2} = K_i,
\end{equation*}
where $C_i$ is the coefficient of $V_i$, $K_i$ is the known term that contains $\rho_i$. We have such an equation for $n-2$ points, excluding the first and $n-1$ one, since $V_0=0$ and $V_{n-1}=$ the bias voltage are known and have to be included in $K_i$'s. The $n-2$ linear equations can be collectively written as
\begin{equation}
  \label{e:cvk} CV=K,
\end{equation}
where $C$ is a $n-2$ by $n-2$ matrix, and $V$ and $K$ are vectors with $n-2$ elements. $C$ is sparse, with at most 3 non-zero elements in each row. It is also symmetric and positive definite. A standard way to solve such a linear equation system is the \emph{conjugate gradient algorithm}~\cite{cg}, which boils down to minimizing a quadratic function of $V$ with the form $V^{T}CV/2-K^T V$. There is a ROOT~\cite{root} macro included in GeFiCa to demonstrate the method. It works well when $n$ is below 100, but becomes painfully slow for a large $n$.

\subsection{Successive Over-Relaxation Method}
\label{s:sor}
Another way to rearrange Eq.~\ref{e:px} is
\begin{equation}\label{e:vic}
  V_{i}=\frac{\frac{\rho}{2\epsilon}+(V_{i+1}/\dd x_+ + V_{i-1}/\dd x_-)/(\dd x_+ + \dd x_-)}
  {(1/\dd x_+ + 1/\dd x_-)/(\dd x_+ + \dd x_-)}.
\end{equation}
If $\rho=0$ and $\dd x_- = \dd x_+$, it can be simplified to
\begin{equation}\label{e:vis}
  V_{i}=(V_{i-1} + V_{i+1})/2,
\end{equation}
where $V_i$ is simply the average of its neighboring values. In both equations, $V_i$ is calculated given $V_{i-1}$ and $V_{i+1}$.

However, since $V_{i-1}$ and $V_{i+1}$ are also unknown (except for $V_0$ and $V_{n-1}$), we need to start the calculation with some initial values. One choice would be $V^{(0)}_0=V^{(0)}_1=\cdots V^{(0)}_{n-2}=0$, and $V^{(0)}_{n-1}=$ the bias voltage, $V_\text{bias}$, where the superscript $^{(0)}$ indicates that these are the initial values of grid points.

Given these initial values, we can use Eq.~\ref{e:vic} or \ref{e:vis} to update $V_i$. Use Eq.~\ref{e:vis} in our calculation hereafter to simplify the demonstration, we have
\begin{equation}
  V_i^{(j)} = [V_{i-1}^{(j-1)}+V_{i+1}^{(j-1)}]/2,
\end{equation}
where $j$ indexes the steps of updating. Since $V^{(j)}_{n-1}=V^{(j-1)}_{n-1}=V_\text{bias}$, it pulls the value of its neighbor $V_{n-2}$ up a bit after each updating, and $V_{n-2}$ pulls up $V_{n-3}$, and so on and so forth. After many iterations, $V_i$ becomes very close to its true value, the difference between the values of $V_i$ in current and previous iteration becomes very small. We can use the following criterion to stop the iteration:
\begin{equation}\label{e:cri}
  \left|\sum_i V_i^{(j)}-\sum_i V_i^{(j-1)}\right| < \text{a small value, e.g.,} 10^{-8}.
\end{equation}
This is the so-called \emph{Successive Relaxation} (SR) method.

\begin{figure}[htbp]\centering
  \includegraphics[width=\linewidth]{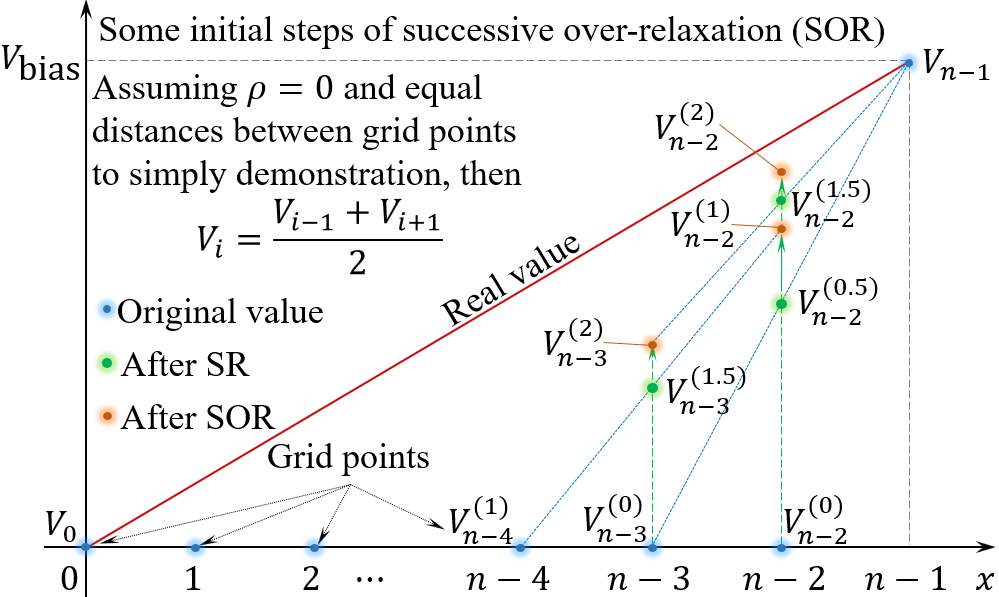}
  \caption{Demonstration of how SOR increases the speed to approach the true value of a potential at a grid point $i$ from an initial guess $V^{(0)}_i$.}\label{f:sor}
\end{figure}

To speed up the relaxation process, we can manually increase the difference of a value between two iterations by introducing a constant, $1 \leq F_R \leq 2$, in the following way:
\begin{equation} \label{e:sor}
  V^{(j)}_i = V^{(j-1)}_i + F_R\times(V^{(j-0.5)}_i-V^{(j-1)}_i),
\end{equation}
where, $V^{(j-0.5)}_i$, is the value updated by the original relaxation method, $F_R$ is called the relaxation factor. This is why the method is called \emph{Successive Over-Relaxation} (SOR) method. The concept of SOR is depicted in Fig.~\ref{f:sor}, where the first a few steps of updating are shown for the last a few grid points in an idea planar detector without any impurity. A carefully chosen relaxation factor can reduce the total number of iteration significantly, which is discussed in detail in Sec.~\ref{s:rf}.

The same method can be applied to multi-dimensional problems in various coordinate systems. The counterparts of Eq.~\ref{e:vic} in those systems are summarized in \ref{a:3dc}.

\subsection{Depletion Voltage}
\label{s:dep}
The general method described in Sec.~\ref{s:adv} applies to numeric calculations as well. We need to search for a $V_\text{d}$ that just depletes the detector with the following procedure:
\begin{enumerate}
  \item Pick up a $V_\text{min}$, say zero, and a $V_\text{max}$, say $10^6$~V.
  \item Assume a bias voltage, $V_\text{bias}\in (V_\text{min},V_\text{max})$.
  \item Run SOR until convergence.
  \item Check if the detector is depleted. If not, replace $V_\text{min}$ with $V_\text{bias}$; if yes, replace $V_\text{max}$ with $V_\text{bias}$.
  \item Repeat from step 2 until $V_\text{max}-V_\text{min}<0.01$~volt.
\end{enumerate}
The depletion voltage $V_\text{d}$ is then $V_\text{bias} \approx V_\text{max} \approx V_\text{min}$ after a successful search.

One feature of the undepleted curve in Fig.~\ref{f:und} distinguishes it from the depleted ones; that is, the maximum or minimum of the potential is not on the boundaries of the detector. Inspired by this, the criterion of depletion in step 4 for numerical calculations can then be set as \emph{none of the grid points has a potential that is larger or smaller than the value of any of its neighboring points}.

A potential drawback of the described method is that it may be time consuming if every new search needs to run an SOR. Fortunately, there is a way to avoid that. As demonstrated in Fig.~\ref{f:und}, the total potential distribution, $V_i$, is a sum of the distribution due to impurity alone, $V_i^\rho$, and the one due to bias alone, $V_i^\text{b}$. Since $V_i^\text{b} = V_i^\text{u} \times V_\text{bias}$, where $V_i^\text{u}$ represents the potential distribution due to unit voltage, 1~V, the total potential distribution can be calculated as
\begin{equation}\label{e:vit}
  V_i = V_i^\text{u} \times V_\text{bias} + V_i^\rho.
\end{equation}
If $V_i^\text{u}$ and $V_i^\rho$ are calculated using SOR before the described iteration, step 3 can be replaced by Eq.~\ref{e:vit} instead of another SOR.

\subsection{Undepleted Region}
\label{s:und}
When $|V_\text{bias}|<|V_\text{d}|$, some region of the detector is not depleted. Numerically, the undepleted region can be found by applying the following procedure to every grid point in the SOR process:
\begin{enumerate}
  \item Calculate the potential of a grid point $V_i$ using potentials of its immediate neighboring points.
  \item Find the maximal and minimal potentials $V_\text{max}$ and $V_\text{min}$ of the immediate neighboring points.
  \item Compare $V_i$ with $V_\text{max}$ and $V_\text{min}$. If $V_i<V_\text{min}$, it is set to be the same as $V_\text{min}$; if $V_i>V_\text{max}$, it is set to be $V_\text{max}$.
\end{enumerate}

\begin{figure}[htbp]\centering
  \includegraphics[width=\linewidth]{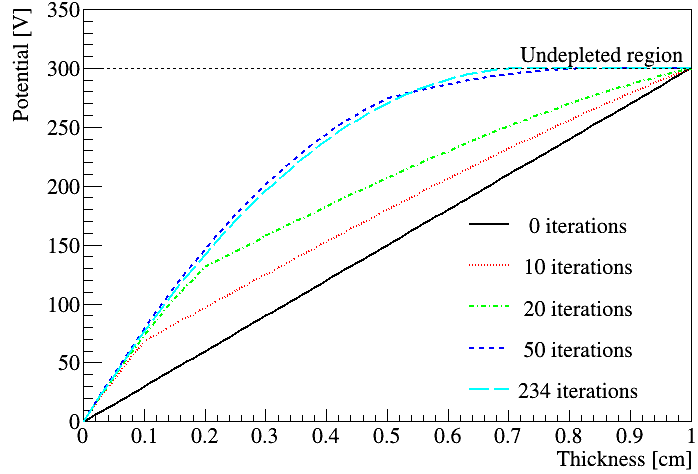}
  \caption{Potential distribution of a planar detector with a $|V_\text{bias}|<|V_\text{d}|$ after some chosen numbers of SOR iterations. It shows the undepleted region as well as the converging process of SOR.}\label{f:conv}
\end{figure}

Fig.~\ref{f:conv} shows potential distributions of a planar detector with a $|V_\text{bias}|<|V_\text{d}|$ after some chosen numbers of SOR iterations. One can see how the undepleted region grows larger near one of the electrodes. Another interesting thing to notice is that it does not take many iterations for the potential to become very close to its final values. Most iterations after that are used to improve the accuracy in a few percent level.

As shown in Fig.~\ref{f:conv}, the undepleted region in a planar detector is adjacent to one of its electrodes.  In the case of a point-contact detector, the undepleted region can stand alone somewhere in the center of the detector, away from any electrode. This is the so-called \emph{pinch-off} effect, since the depleted region is ``pinched off'' from electrodes.

Since there is no electric field in the undepleted region to separate and drift electrons and holes generated by radiation interactions, the region is insensitive to radiation.  Even if a pair of charge carriers are generated in the depleted region, one of them may drift to the undepleted region along the electric field and get stuck there instead of being collected by an electrode. It is hence worthwhile to reveal the existence of such an undepleted region through field calculation.

\begin{figure}[htbp]\centering
    \includegraphics[width=\linewidth]{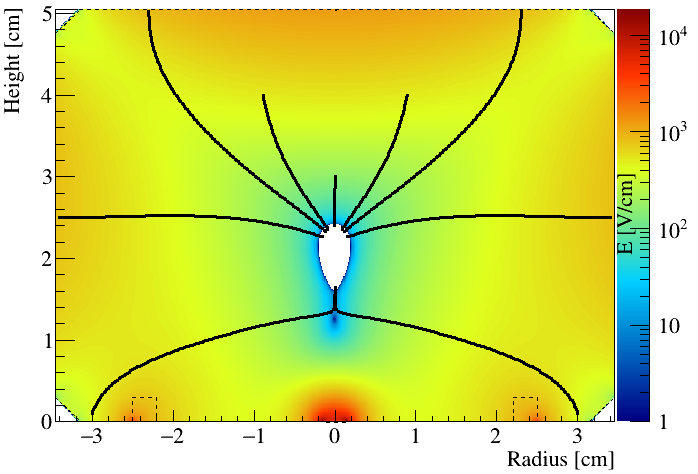}
    \caption{The \emph{pinch-off} effect demonstrated by the electric field as a colored contour in logarithm scale in a point-contact detector.}
    \label{f:pof}
\end{figure}

Fig.~\ref{f:pof} shows the electric field as a colored contour in logarithmic scale in a point-contact detector. The point contact is at the origin of the plot. The field value around the center of the detector is very close to zero, so the logarithm of them approaches negative infinity. They are color coded as white, which nicely visualizes the undepleted region that is \emph{pinched off} from the point contact.

\begin{figure}[htbp]\centering
    \includegraphics[width=\linewidth]{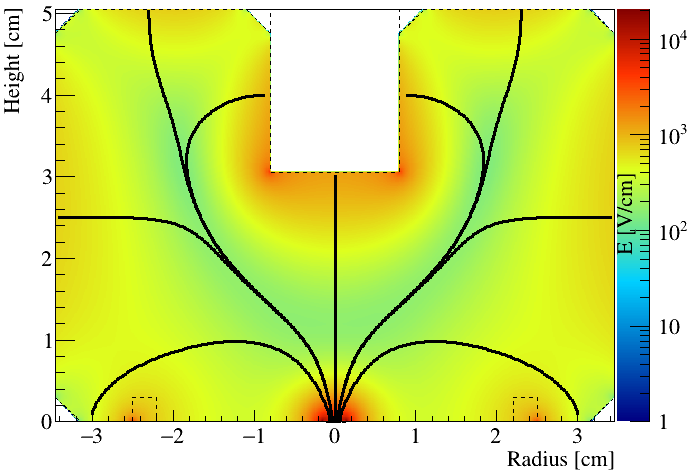}
    \caption{Electric field distribution in an ICPC.}
    \label{f:icpc}
\end{figure}

This phenomenon seriously limits the size of a point-contact detector, since the electric field inside the detector becomes weaker when the size of the crystal becomes larger if the bias is not ramped up accordingly. To avoid this, one can bore a central hole from the opposite side of the point-contact, metallize its surface and keep it at the same bias as other surfaces. Such a detector is called a inverted-coaxial point-contact detector, or ICPC in short~\cite{icpc}. Fig.~\ref{f:icpc} shows the electric field distribution in an ICPC as a color coded contour in logarithmic scale. Other configurations of this calculation are kept the same as the ones used to generate Fig.~\ref{f:pof}, including the crystal impurity level and the bias voltage. One can see clearly that the undepleted region is successfully eliminated from the center of the crystal.

\subsection{Electric Field Lines}
\label{s:fl}
The thick black lines in Fig.~\ref{f:pof} and \ref{f:icpc} are estimated charge drift trajectories starting near the outer surface of the detectors. The procedure of the estimation can be illustrated in two dimensional Cartesian coordinates:
\begin{enumerate}
  \item Linearly interpolate the electric field components $E_x, E_y$ at a random starting point $(x, y)$ using values at its four neighboring grid points.
  \item Calculate the total electric field $E=\sqrt{E_x^2+E_y^2}$ at the same point.
  \item Calculate the propagation of a positive unit charge along $x$ and $y$: $\dd{x} = \mu E_x \dd{t}, \dd{y} = \mu E_y \dd{t}$, where $\mu$ takes a value of $40,000$~cm$^2$/(volt$\cdot$second), a number in between typical electron and hole drift mobilities in HPGe crystals~\cite{mihailescu00, reggiani78, bruyneel06}, and $\dd{t}$ takes a value of 10~ns.
  \item $\dd{x}$, $\dd{y}$ are then further modified using equations $\dd{x}=\dd{x}\times weight, \dd{y}=\dd{y}\times weight$, where $weight$ = (5~volt/mm)/$E$, which is used to stretch $\dd{x}, \dd{y}$ in a weak field and shrink them in a stronger one.
  \item The new position of the positive unit charge is then updated to $(x+\dd{x}, y+\dd{y}$), which is saved in an object of the \emph{FieldLine} class.
  \item Repeat step 1 to 5 using the updated starting point coordinates until it moves out of the crystal or falls into an undepleted region.
\end{enumerate}
Changing the positive unit charge to a negative one would let it propagate to the opposite direction.

Ignoring the influence of the germanium crystal structure, charge carriers drift roughly along electric field lines. The propagation path created this way can hence be regarded as a rough estimation of the field line.

It is interesting to see in Fig.~\ref{f:icpc} that the field lines merge in the middle of the detector and get collected at the point contact, just as streams flow down to a river in a valley (the blue-ish region in Fig.~\ref{f:icpc} if color-printed).

\subsection{Boundaries in between Grid Points}
\label{s:bcp}
\begin{figure}[htbp]\centering
  \includegraphics[width=0.9\linewidth]{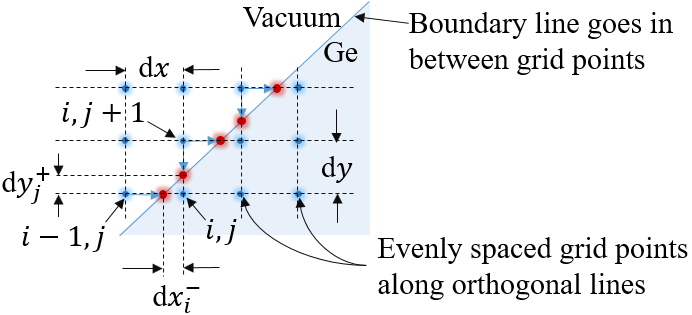}
  \caption{Variable distances between grid points in case of a boundary line goes in between.}\label{f:tl}
\end{figure}

Sometimes, the edges between the side and end surfaces of a cylindrical detector are tapered, shown as the small white triangular regions at the corners of the color contours in Fig.~\ref{f:pof} and \ref{f:icpc}.  A crystal boundary line hence can go in between grid points that are distributed along orthogonal lines, as shown in Fig.~\ref{f:tl}. Assuming a simple case, where the grid points are evenly spaced, the distances between them take fixed values, $\dd{x}$ and $\dd{y}$. The distances of a regular grid point to its previous and next neighbors equal to each other: $\dd{x}_-=\dd{x}_+=\dd{x}$. In case of a point near the boundary, such as $(i,j)$ shown in Fig.~\ref{f:tl}, it is more precise to replace $\dd{x}_-$ with $\dd{x}_i^-$ when evaluating Eq.~\ref{e:ndv} along the $x$-axis, where $\dd{x}_i^-$ is the distance to the boundary instead of the distance to the previous grid point as shown in Fig.~\ref{f:tl}. Similarly,  $\dd{y}_+$ should be replaced by $\dd{y}_i^+$ for a more precise evaluation of Eq.~\ref{e:ndv} along the $y$-axis. This effectively moves nearby points on the vacuum side right to be on the boundary, shown as the red dots in Fig.~\ref{f:tl}. Such an operation can only be done if variable step lengths are allowed for individual grid points.

\subsection{Weighting Potential in Segmented Detectors}
\begin{figure}[htbp]\centering
    \includegraphics[width=\linewidth]{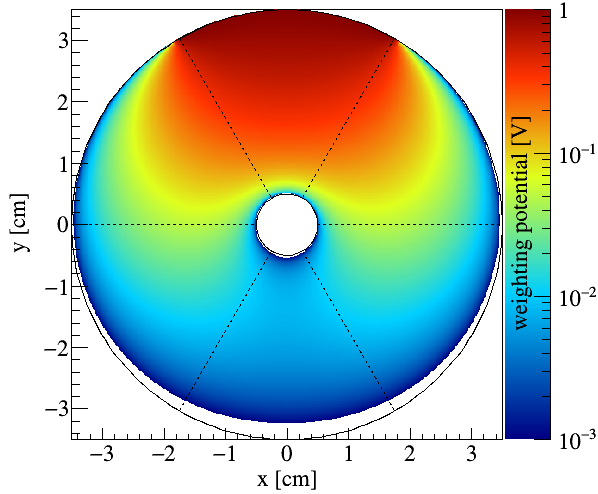}
    \caption{Weighting potential distribution of a segmented detector (six evenly distributed segments along the azimuthal direction).}
    \label{f:sip}
\end{figure}

In addition to the real electric field, the so-called \emph{weighting potential} is also of great interest, since it can be used to calculate the electric charges on an electrode induced by the drifting charge carriers inside a detector based on the Shockley-Ramo's Theorem~\cite{he01, radeka88, gatti82}. It differs from a real potential in two ways. First, it is purely determined by its  boundary conditions. The impurity concentration in a crystal should be regarded as zero in calculating the weighting potential. Second, the voltage on the interested electrode should be set to 1~volt, while the voltage on any other electrode should be set to zero in calculating the weighting potential. The weighting potential of an electrode is probably best demonstrated using a segmented HPGe detector. Fig.~\ref{f:sip} shows the cross section of a detector segmented evenly in six along the azimuthal direction. The weighting potential of one of the segment electrode is overlaid as a colored contour in a logarithm scale. The white circle in the middle indicates the core electrode of this cylindrical detector. The colored contour does not quit reach the bottom boundary, simply because the potential there is too close to zero to be color coded in a logarithm scale.

\subsection{Capacitance}
The capacitance of a HPGe detector $C_d$ is of special interest due to at least two reasons. First, the electronic noise of a HPGe detector is proportional to the sum of $C_d$ and the capacitance of the feedback capacitor, $C_f$, in the pre-amplifier circuit of the detector~\cite{radeka88, nashashibi90, nashashibi92}. Second, $C_d$ decreases as the detector bias voltage ramps up. The reason becomes clear later in this section. This feature can be used to measure the depletion voltage, $V_\text{d}$. It can also be used to check if a detector operates normally during the ramping up of its bias voltage. It is therefore an important task of a field calculation package to calculate $C_d$ given an arbitrary bias voltage.

\begin{figure}[htbp]\centering
    \includegraphics[width=\linewidth]{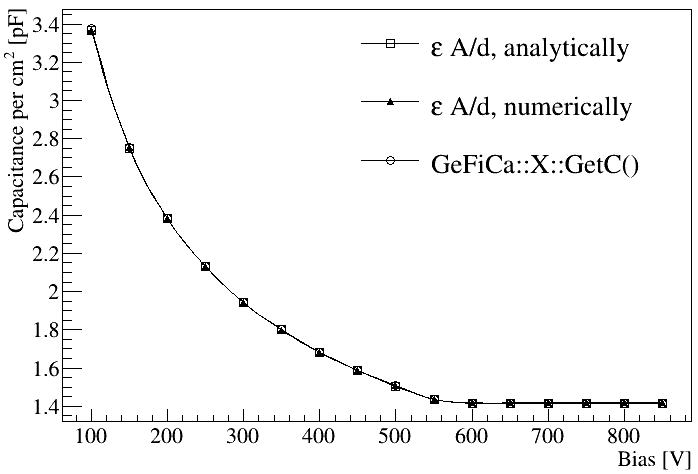}
    \caption{Capacitance per unit area versus bias voltage of an ideal planar detector calculated in three different ways as detailed in the text. The agreement between each other verifies the correctness of the numerical calculation of the detector capacitance.}
    \label{f:cv}
\end{figure}

For an ideal one dimensional planar detector,
\begin{equation}\label{e:cd}
C_d=\epsilon A/d,
\end{equation}
where $A$ is the area of an electrode of the detector and $d$ is the thickness of the depleted region in the detector, which can be calculated as
\begin{equation}\label{e:dvd}
  d=\sqrt{2\epsilon V_\text{d}/\rho}.
\end{equation}
This relation can be derived from Eq.~\ref{e:asp} with the boundary condition \ref{e:b1} and \ref{e:b2}. $C_d$ hence is anti-proportional to $\sqrt{V_\text{d}}$, and decreases as $V_\text{d}$ increases, until $d$ becomes the thickness of the plan detector. After that, $C_d$ stays at its minimum since $d$ cannot increase anymore. The square data points in Fig.~\ref{f:cv} are calculated using Eq.~\ref{e:cd} and \ref{e:dvd} given individual bias values.

The depletion depth $d$ can also be determined numerically using the method described in the previous section. $C_d$ can be then calculated using Eq.~\ref{e:cd}. The results are the triangle data points in Fig.~\ref{f:cv}.

For a detector configuration as complex as a point-contact one, there is no analytic solution for $C_d$. The following numerical method is used in GeFiCa to calculate $C_d$. It is based on the fact that the energy stored in a charged capacitor $U$ is equal to the overall work done $W$ to move a total amount of charges $Q$ to the electrodes against the electric field $E$ caused by $Q$ stored in the capacitor:
\begin{equation}
  U=W. \label{e:uew}
\end{equation}
Given an arbitrary amount of charges $q$ already stored in a capacitor, the work done to increase it by an infinitesimal amount $\dd{q}$ is
\begin{equation}
  \dd{W} = V_\text{bias} \dd{q} = (q/C_d)\dd{q}.
\end{equation}
Integrating it on both sides yields
\begin{equation} \label{e:EnergyinCap}
  W=\int_{0}^{Q}\frac{q}{C_d} dq=\frac{1}{C_d}\int_{0}^{Q}q dq=\frac{Q^2}{2C_d}=\frac{C_dV_\text{bias}^2}{2}.
\end{equation}
The relation $C_d=Q/V_\text{bias}$ is used in the last step of the derivation to replace $Q$, an unknown variable, with $C_d$ and $V_\text{bias}$.

On the other hand, since the electric field energy density is $\epsilon E^2/2$, $U$ can be expressed as
\begin{equation} \label{e:EnergyinElec}
  U=\frac{1}{2}\epsilon\int_{V}E^2 \dd{\tau},
\end{equation}
where $\dd{\tau}$ is the volume integration element. For a planar detector with a constant impurity, the integral can be solved analytically as:
\begin{equation}\label{e:up}
  U=\frac{1}{2}\epsilon E^2\int_{V} \dd{\tau}=\epsilon V^2_\text{bias}A/(2d).
\end{equation}
Replacing $U$ and $W$ in Eq.~\ref{e:uew} with Eq.~\ref{e:up} and \ref{e:EnergyinCap}, we derive Eq.~\ref{e:cd}.

The numerical version of Eq.~\ref{e:EnergyinElec} for an ideal planar detector in Cartesian coordinates is
\begin{equation} \label{e:un}
  U\approx\frac{1}{2}\epsilon\sum_{i=0}^{n}E_i^2 \dd{x_i}A,
\end{equation}
where $i$ is the index of each grid point. Combining Eq.~\ref{e:un}, \ref{e:EnergyinCap} and \ref{e:uew}, $C_d$ per unit area $A$ can be calculated as
\begin{equation} \label{e:cdn}
  C_d/A = \epsilon \sum_{i=0}^{n}E_i^2 \dd{x_i}/V^2_\text{bias}.
\end{equation}
This is implemented in function \emph{GeFiCa::X::GetC()}. The results are shown as the circle data points in Fig.~\ref{f:cv}. The perfect agreement between all methods verifies two numerical calculations in GeFiCa: the finding of the undepleted region (or depleted region) and the calculation of $C_d$ given an arbitrary $V_\text{bias}$.

It is worth noting that the electric field $E$ here is only due to $Q$ accumulated on the detector electrodes. It is different from the actual field in a depleted detector which is the combination of the fields from both $Q$ and the space charge in the crystal.

For a point-contact detector in Cylindrical coordinates, the numerical version of Eq.~\ref{e:EnergyinElec} is
\begin{equation}\label{e:uppc}
  U\approx \frac{1}{2}\epsilon\sum_{i=0}^{n}E_i^2r_i\dd{r_i}\dd{z_i}\int_{0}^{2\pi}\dd{\theta}.
\end{equation}
It is implemented in function \emph{RhoZ::GetC()}.

\subsection{Interpolation Between Grid Points}
\begin{figure}[htbp]
  \centering
  \includegraphics[width=0.65\linewidth]{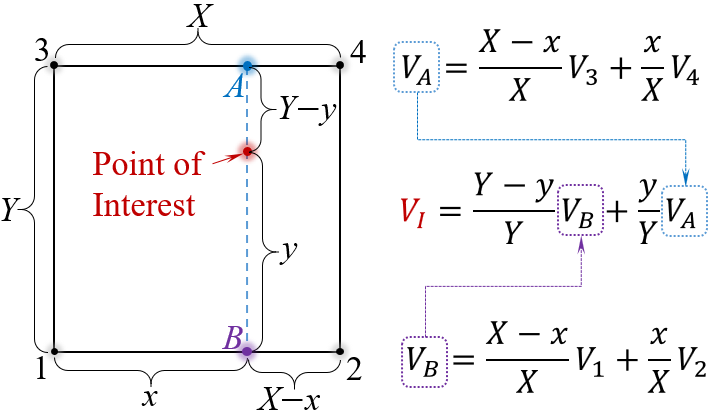}
  \caption{Rectangular interpolation of the potential at the point of interest, $V_I$, using potentials at its nearest grid points, $V_1,V_2,V_3$ and $V_4$ in a 2D Cartesian grid.}
  \label{f:int}
\end{figure}

A numeric calculation can only give the field values right at each grid point. Interpolations are needed to get the field values at a random point that may not coincide with any grid point. Fig.~\ref{f:int} shows the equations to linearly interpolate the potential value at the point of interest that falls in between four points in a 2D Cartesian grid using the known potential values on those four points, $V_1, V_2, V_3$ and $V_4$, taking into account the distances between points, $X,Y,x,y$.

\begin{figure}[htbp]
  \centering
  \includegraphics[width=\linewidth]{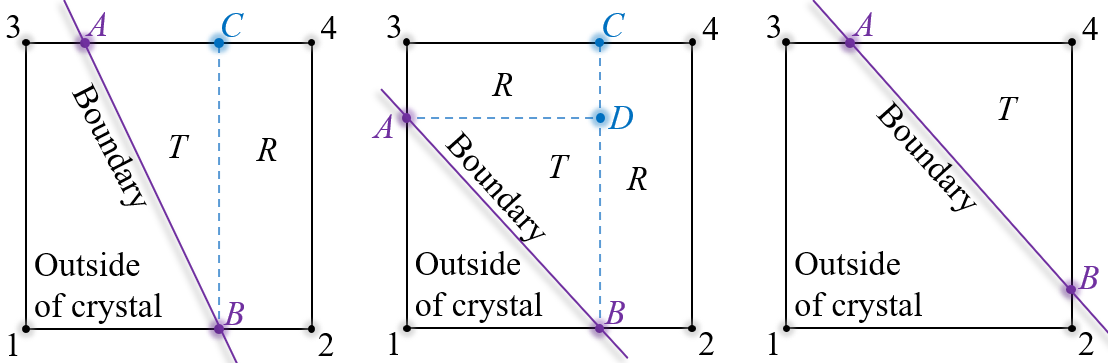}
  \caption{Three ways that a boundary line goes through a unit grid square. Within the crystal, the point of interest can fall into either a triangular or a rectangular region marked as \emph{T} or \emph{R}.}
  \label{f:b3}
\end{figure}

\begin{figure}[htbp]
  \centering
  \includegraphics[width=0.65\linewidth]{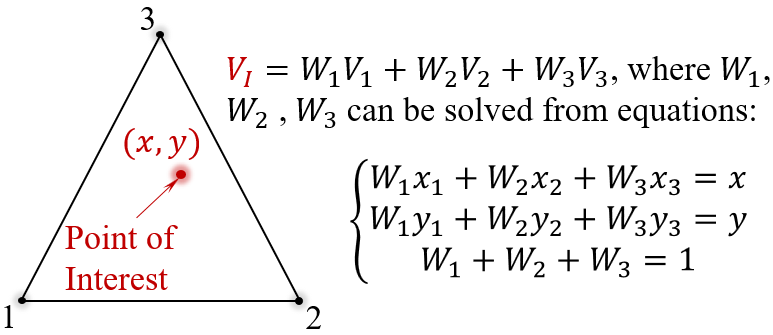}
  \caption{Triangular interpolation of the potential at the point of interest, $V_I$, using potentials at its nearest 3 grid points, $V_1,V_2,V_3$.}
  \label{f:bary}
\end{figure}

This method does not work for unit grid squares across crystal boundaries that are neither in parallel with nor perpendicular to grid lines, since those boundary lines can separate a square into irregular shapes, the interpolation of which can be complicated. There are three ways that such a boundary line can go through a unit grid square as shown in Fig.~\ref{f:b3}. Potentials at the crossing point, $V_A$ and $V_B$, are equal to the bias applied to that boundary. Most of the time the field outside of the crystal is not of interest. Within the crystal, the point of interest can fall into either a triangular or a rectangular region marked as \emph{T} or \emph{R}, separated by blue dotted lines in Fig.~\ref{f:b3}. If it falls into an \emph{R} region, the interpolation method shown in Fig.~\ref{f:int} can be used. If it falls into a \emph{T} region, the triangular interpolation shown in Fig.~\ref{f:bary} can be used.

The potential at the point of interest, $V_I$, in a \emph{T} region can be calculated as the weighted sum of potentials at the grid points around, $V_1, V_2$ and $V_3$. The weights, $W_1, W_2$ and $W_3$ are the coordinates of the point of interest in the barycentric coordinates defined by the three grid points around. Since the Cartesian coordinates of all grid points and the point of interest are known, $W_1, W_2$ and $W_3$ can be calculated by transforming the Cartesian coordinates of the point of interest to the barycentric coordinates:
\begin{align*}
  W_1 &= \frac{(y_2-y_3)(x-x_3)+(x_3-x_2)(y-y_3)}{(y_2-y_3)(x_1-x_3)+(x_3-x_2)(y_1-y_3)},\\
  W_2 &= \frac{(y_3-y_1)(x-x_3)+(x_1-x_3)(y-y_3)}{(y_2-y_3)(x_1-x_3)+(x_3-x_2)(y_1-y_3)},\\
  W_3 &= 1 - W_1 - W_2,
\end{align*}
where $(x_1, y_1)$, $(x_2, y_2)$ and $(x_3, y_3)$ are the Cartesian coordinates of the grid points 1, 2 and 3 shown in Fig.~\ref{f:bary}, $(x, y)$ are the Cartesian coordinates of the point of interest.

In case of the vector field $E$, interpolations are done separately for individual components to get $E_x, E_y$ at the point of interest. The total $E$ is then calculated as $\sqrt{E_x^2 +E_y^2}$.

\section{Implementation}
\subsection{Coding Convention}
The coding convention is similar to that of ROOT~\cite{convention}. For example,
\begin{itemize}
  \item Classes and functions all start with capital letters. Word boundaries are indicated by \textit{CamelCase}.
  \item Classes names are all nouns.
  \item Function names are all verbs.
  \item Private member variables all start with letter \textit{f}.
  \item Boolean variables/functions start with \emph{Is/Are}.
  \item Indentation is made by three spaces instead of a hard tab to ensure the same appearance of the codes in different editors.
\end{itemize}
The following exceptions are used to increase the readability of the codes to the user:
\begin{itemize}
  \item Class names do not have prefix letters, such as \textit{T} in ROOT. Instead, the name space \textit{GeFiCa} is used to avoid name collision should GeFiCa be used together with other libraries.
  \item Configurable member variables are made public to avoid trivial getters and setters. Their first letters are capitalized. Unlike private member variables, they do not have letter \textit{f} prefixed.
\end{itemize}

\subsection{Class Structure}
\begin{figure}[htbp]\centering
  \includegraphics[width=0.75\linewidth]{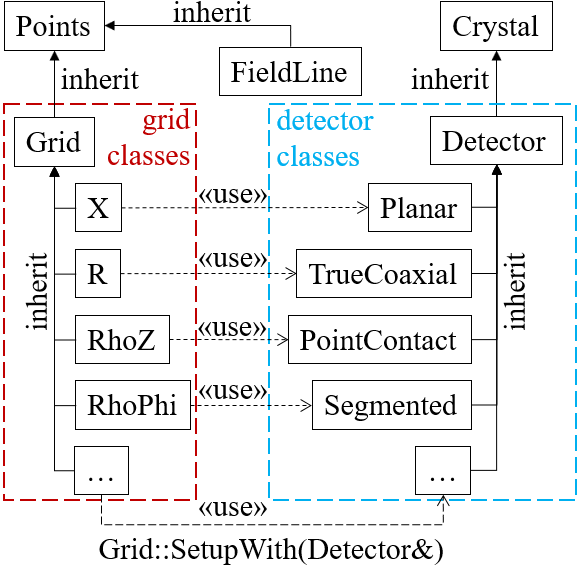}
  \caption{Relation between GeFiCa classes.}
  \label{f:flow}
\end{figure}

As shown in Fig.~\ref{f:flow}, most of the GeFiCa classes belong to two categories: \emph{grid} and \emph{detector}. Those that are derived from class \emph{Grid} are used to describe grid setups. Those that are derived from class \emph{Detector} are used to describe detector configurations. The \emph{Grid} class inherits a set of arrays from the \emph{Points} class to describe variables associated with individual grid points, such as coordinates and field values. Names of its derived classes indicate the dimension and coordinates used to construct the grid. For example, \emph{X} is used for a one dimensional grid in Cartesian coordinates, \emph{RhoZ} is used for a two dimensional grid in cylindrical coordinates. The \emph{Detector} class inherits impurity setup from the \emph{Crystal} class. Its derived classes, such as \emph{PointContact}, inherit from it the common detector setups, such as bias voltages. A grid class can get boundary conditions and the impurity distribution from a corresponding detector class through a virtual function interface defined in the \emph{Grid} class:
\begin{lstlisting}[language=c++]
virtual void Grid::SetupWith(Detector&);
\end{lstlisting}
This is demonstrated in the following code snippet:
\begin{lstlisting}[language=c++]
RhoZ grid; //create 2D Cylindrical grid
PointContact detector; //create detector
//setup grid with detector configuration
grid.SetupWith(detector);
\end{lstlisting}

The data flow can be the other way around, that is, a detector class gets grid setups from a grid class. However, since it is the grid that the SOR process updates instead of the detector configuration, this is a less natural choice. With the current data flow direction, the SOR can then be performed by simply calling
\begin{lstlisting}[language=c++]
  grid.SuccessiveOverRelax();
\end{lstlisting}

Another choice would be to combine the detector and grid classes. For example, instead of having both \emph{PointContact} and \emph{RhoZ}, we can create a single class called \emph{PointContactRhoZ}. The advantage of this approaches is that there is no need to pass information from the latter to the former through some interface functions. The drawback is the lack of clarity, the same class object will be used for both detector configuration and grid operation. Considering the main purpose of GeFiCa is to demonstrate the logic, methods, and techniques for field calculation, we chose not to use this approach.

To its root, this is actually a question of to what extend we want to utilize the object-oriented (OO) coding style. Think about two extreme cases. First, we can write everything in a single main function. Second, we can create a class for each individual functionality, such as the impurity profile and the bias voltage. The first approach relies on careful documentation to clarify its internal logic. The second introduces many trivial interfaces to pass information between classes. A balanced approach in between is adopted for GeFiCa.

\subsection{Data Structure and I/O}
\label{s:dsio}
Minimally, two float numbers are needed for each point in a one dimensional grid with a fixed interval between its points: one for the spacial coordinate and another for the electric field potential. The number of points in a grid must be changeable according to the dimension of a detector and the precision of a calculation. This demands the use of arrays that can change in size to store variables of individual points. Even though a \emph{float} number is precise enough to hold the final result of a numerical calculation, a \emph{double} is preferred to preserve precision during iterations of a SOR process. A standard C++ \emph{vector<double>} is used in GeFiCa for each variable to provide enough precision and flexibility.

Given a grid with variable step sizes as shown in Fig.~\ref{f:grid}, one more variable is needed for each point to store the distance to its next neighbor, $\dd{x}_i^+$. The distance to its previous neighbor is saved as $\dd{x}_{i-1}^+$ in its previous point. In GeFiCa, however, the distances to both the previous and next neighbors, $\dd{x}^-$ and $\dd{x}^+$, are saved, since they may be different for some of the boundary points as detailed in Sec.~\ref{s:bcp}. This certainly creates redundancy in storage for points away from boundaries. However, since output files of GeFiCa are saved in ROOT format instead of plain ASCII, such redundancy does not increase their sizes much due to the \emph{gzip} compression algorithm used in ROOT to save equal-value variables only once.

In principle, electric field values can be calculated from the potential using Eq.~\ref{e:edv} when needed. However, given their frequent usage, their values on each point are calculated and saved in GeFiCa after the SOR calculation for the potential.

For a three dimensional grid with a variable step size, 14 \emph{std::vector}s with a \emph{double} precision are needed in total to save 3 coordinates, $3\times2$ distances to previous and next points, 1 potential, 1 total electric field and its 3 components. They are public member variables in the class \emph{Points} inherited by all \emph{grid} classes shown in Fig.~\ref{f:flow}, including those representing lower dimensional grids, where variables for higher dimensions are not used at all. Since the C++ \emph{vector} does not allocate memory if it has no element, there is no penalty in storage size in this solution. An alternative is to create \emph{Points1D}, \emph{Points2D}, \emph{Point3D}, and consequently, \emph{Grid1D}, \emph{Grid2D}, \emph{Grid3D} for various dimensions. This complicates the overall class structure unnecessarily, hence is not used in GeFiCa.

A few more \emph{vector}s are added in the \emph{Grid} class to record space charge densities in individual grid points, as well as flags to tell whether a point is in or out of a crystal, and whether it is in or out of the depleted region.

As described in Section~\ref{s:fl}, an electric field line can be saved in a series of points with variable distances between them. That is why the class \emph{FieldLine} inherits the data structure from \emph{Points}, as shown in Fig.~\ref{f:flow}.

\begin{figure}[htpb]\centering
  \includegraphics[width=\linewidth]{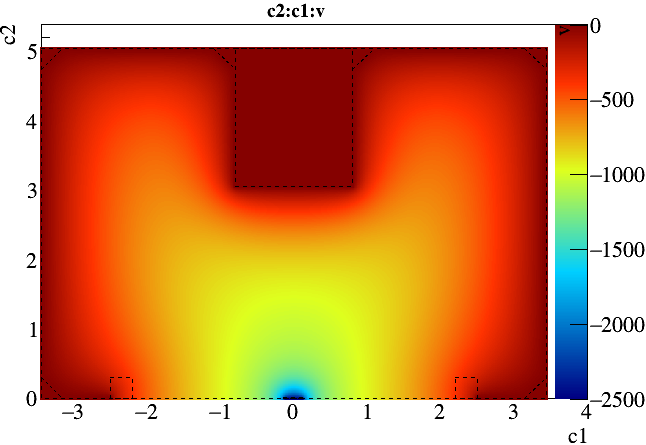}
  \caption{The color contour of the potential field of an ICPC detector drawn with \emph{TTree::Draw("c2:c1:v","","colz")}.}
  \label{f:v}
\end{figure}

The \emph{Grid} and the \emph{Detector} classes are both daughters of the \emph{TNamed} class in ROOT, which inherits the capability to stream its data members for I/O from the \emph{TObject} class in ROOT. Consequently, all concrete grid and detector classes can be directly saved into a standard ROOT file using one line of C++ as shown in the following code snippet:
\begin{lstlisting}[language=c++]
  TFile file("ICPC.root","recreate");
  detector.Write(); // save config.
  grid.Write(); // save grid
  file.Close();
\end{lstlisting}
As described previously, repeated numbers in the ROOT file are compressed to save storage space. Detailed benchmark of the file size can be found in Section~\ref{s:fs}. After opening the ROOT file in a ROOT interactive session, one can use the Cling meta command \emph{.ls} to list the saved objects:
\begin{lstlisting}[language=sh]
root ICPC.root
root[].ls
TFile** ICPC.root
TFile*  ICPC.root
KEY: GeFiCa::PointContact pc;1 detector
KEY: GeFiCa::RhoZ rhoz;1 2D grid...
\end{lstlisting}
and directly use the loaded objects (\emph{rhoz} and \emph{pc}) to investigate and visualize the field and the detector:
\begin{lstlisting}[language=c++]
root[] TTree *t = rhoz->GetTree()
root[] t->Draw("c2:c1:v","","colz")
root[] pc->Draw()
\end{lstlisting}
The first line creates a \emph{TTree} object \emph{t} out of the saved field values in the \emph{rhoz} object. The second line draws the potential, \emph{v}, on the first ($y$-axis) and second ($x$-axis) coordinates, \emph{c1} and \emph{c2}, as a colored contour along $z$-axis (the "colz" option), as shown in Fig.~\ref{f:v}.

The price to pay for all these convenience is that the objects saved in the ROOT file can only be loaded without warning message when the compiled GeFiCa library can be found and automatically loaded by ROOT. The way to realize this is detailed in Section.~\ref{s:ci}.

\subsection{Detector Configurations}
Two pieces of information are needed for electric field calculation: first, boundary conditions, and second, the space charge distribution.

\begin{figure}[htbp]
  \centering
  \includegraphics[width=\linewidth]{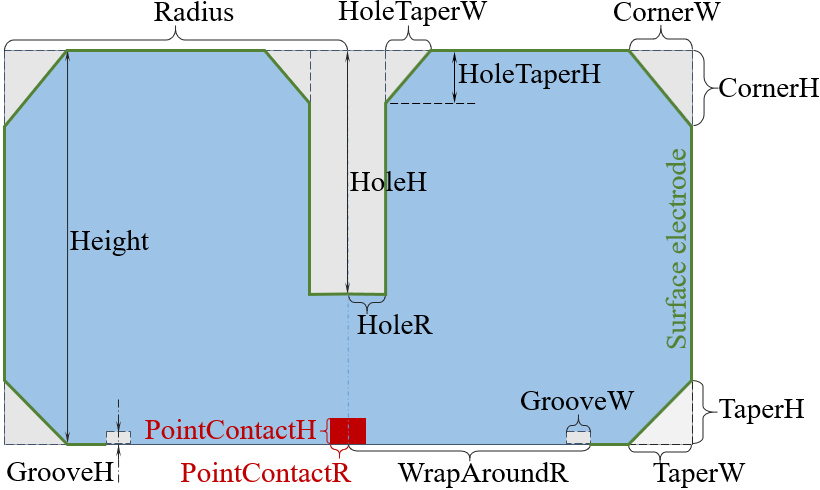}
  \caption{Cross section along $z$-axis of an inverted coaxial point-contact HPGe detector, and parameters describing its dimensions. To shorten the names, width is represented as a single capital case ``W'', height ``H'', and radius ``R''.}
  \label{f:pars}
\end{figure}

Boundary conditions can be set through the detector geometry and voltages on electrodes. Take the previously defined \emph{PointContact} detector as an example, its basic dimensions can be set as
\begin{lstlisting}[language=c++]
  detector.Radius=3.45*cm;
  detector.Height=5.05*cm;
  detector.PointContactR=1.4*mm;
  detector.PointContactH=0.1*mm;
\end{lstlisting}
A full list of geometry parameters that can be set for a \emph{PointContact} detector is shown in Fig.~\ref{f:pars}. Its bias voltages can be set as an array:
\begin{lstlisting}[language=c++]
  // point-contact voltage
  detector.Bias[0] = - 2.5*kV;
  // surface contact voltage
  detector.Bias[1] = 0*volt;
\end{lstlisting}
In case of a segmented detector, the bias voltage array can have more than two elements. The index of an element can be kept the same as the corresponding segment identification number.

As described in detail in Sec.~\ref{s:sc}, it is reasonable to use a first-order polynomial to approximate the space charge distribution in a HPGe crystal. With this simplification, we just need to specify the impurities at the top and the bottom of a crystal given by the manufacturer. For example,
\begin{lstlisting}[language=c++]
  detector.BottomImpurity=3e9/cm3;
  detector.TopImpurity=7e9/cm3;
\end{lstlisting}
The impurity level at a specific axial position is interpolated in GeFiCa based on these two numbers.
In case of a small crystal, the impurity can be regarded as a constant. Its average impurity can be set as
\begin{lstlisting}[language=c++]
  detector.SetAverageImpurity(3e9/cm3);
\end{lstlisting}

\subsection{Units and Constants}
We have seen in previous code snippets that an input parameter in GeFiCa is a product of a number and a unit. Common units and constants for field calculation, together with their conversion rules, are defined in GeFiCa/src/Units.h. The following is a snippet of the file:
\begin{lstlisting}[language=c++]
namespace GeFiCa {
  static const double C=1; // Coulomb
  static const double cm=1;
  static const double cm3=cm*cm*cm;
  static const double mm=0.1*cm;
  static const double volt=1;
  static const double kV=1000*volt;
  // vacuum permittivity [C/volt/cm]
  static const double epsilon0
            = 8.854187817e-14*C/volt/cm;
  // dielectric constant of Ge
  static const double epsilonR=16;
}
\end{lstlisting}
The advantage of this unit system is three-fold. First, the code is self-explainable, there is no ambiguity in the unit of an input value. Second, the user has freedom to choose units, such as ``mm'' instead of ``cm'', or ``kV'' instead of ``volt''. Otherwise, he or she has to use the set of units used for internal calculation. Third, since the unit conversion rules are pre-defined, there is no need to worry about them when using input parameters for internal calculations. The programmer can focus on the logic instead of unit conversion.  This way of handling units is adopted from Geant4~\cite{g41, g42, g43}.  Most of the units and constants have been defined in Geant4 already. However, since only a small subset of the units are useful for field calculation, they are re-defined in GeFiCa to avoid unnecessary dependence on Geant4.

\subsection{Compilation and Installation}
\label{s:ci}
GeFiCa relies on ROOT to realize C++ and Python scripting, efficient I/O and plotting. It has to be compiled against ROOT libraries. This is achieved through a simple Makefile that uses the \lstinline{root-config} executable available from any successfully installed ROOT package to get the location of ROOT libraries and necessary compilation flags. The compilation process is as simple as

\begin{lstlisting}[language=bash]
  cd /path/to/GeFiCa/src && make
\end{lstlisting}

After a successful compilation a shared C++ library, \lstinline{libGeFiCa.so}, can be found in  \lstinline{/path/to/GeFiCa/src}.
Once its location is added to the \lstinline{LD_LIBRARY_PATH} environment variable (or \lstinline{DYLD_LIBRARY_PATH} in MacOS), the library can be automatically loaded only when needed as any other ROOT libraries in ROOT interactive sessions or scripts thanks to the \emph{rootmap} and \emph{pcm} files~\cite{rmap} generated by the \emph{make} process in the same directory as the library.

\subsection{Supported OS}
Since ROOT is available in the three common operating systems, Linux, Windows and MacOS, in principle, GeFiCa should be able to be compiled in all of them as well. However, since GeFiCa relies on a simple Makefile to compile, it cannot be directly compiled through the native Windows compilation system. Instead, it can be compiled in  a Windows Subsystem for Linux (WSL). To date, GeFiCa has been compiled successfully in CentOS 6 and 7, MacOS 10.12, 10.13, 10.14, and Ubuntu 18.04 as a WSL.

\subsection{Code Accessibility}
The codes of GeFiCa are hosted online at GitHub~\cite{github}: \url{https://github.com/jintonic/gefica}. They can be downloaded directly from the web page or through \emph{git}.
GeFiCa is release under the MIT License~\cite{mit}. It can be freely used without any warranty as long as the license is distributed along with it.

\subsection{Code Documentation}
A git branch \emph{gh-pages} is used to host the homepage code for GeFiCa. The homepage is available under a customized domain name: \url{http://physino.xyz/gefica}. It lists three main resources about GeFiCa that one can get help from: the GeFiCa repository page hosted on GitHub, the code documentation hosted on \url{https://codedocs.xyz}, and the user manual hosted on \url{https://readthedocs.org}.

There is a \emph{README.md} file in each directory in GeFiCa to explain the contents of the directory written in \emph{GitHub Flavored Markdown} format~\cite{md}. They are rendered to web pages automatically in GitHub. A user can quickly get help with or without the source code.

Explanations of GeFiCa classes and variables are embedded in the source code as C++ comments using the Doxygen~\cite{doxygen} convention. They can be rendered by Doxygen into nicely formatted documentations locally or on \url{https://codedocs.xyz}. The online version is updated automatically once new codes are pushed to the GitHub repository.

The user manual is written in \emph{restructured text} format and can be rendered to web pages locally or on \url{https://readthedocs.org}. The online version is updated automatically once new documentation is pushed to the GitHub repository.

In addition to these, example codes are shipped with GeFiCa as ROOT macros as described in detail in the next section to demonstrate the usage of individual GeFiCa classes.

\subsection{Macros and Scripts}
\label{s:ms}
A modern C++ interpreter, cling~\cite{cling}, has been created and adopted as the back-end of the interactive session of ROOT~\cite{root} since the version 6 of it. A user can run C++ snippets, sometimes called ROOT macros or scripts, interactively in cling without writing and compiling the ``main'' function. With immediate feedback after the execution of each line of a script, a user can learn and experiment a new C++ class, a function, or simply a syntax easily. To fulfill its educational purpose, GeFiCa is compiled as a ROOT library. All snippets in previous sections demonstrating the configuration of a detector or the operation of a grid can be run as they are in cling.

ROOT also provides a Python extension module, PyROOT, that allows the user to interact with any ROOT class from the Python interpreter. For users who prefer the Python interpreter to cling, they can call GeFiCa classes with Python syntax directly in the standard Python interpreter.

It is worth noting that cling comes with a Jupyter~\cite{jupyter} kernel, which makes it possible to run GeFiCa scripts in a Jupyter notebook with either C++ or python syntax.

All concrete grid and detector classes in GeFiCa inherit the capability to inspect themselves from the \emph{TObject} class in ROOT. Some standard functions in TObject, such as \emph{Dump()}, can be used to check the default or user-specified configurations of a grid or detector object, as shown in Listing~\ref{l:x}. The first column of the output are the member variables of the \emph{GeFiCa::X} class. The second are their current values. The last are explanations of those variable. These explanations are written as C++ comments after the member variables. They can be parsed by both Doxygen and ROOT to generate code documentation in various formats and contexts.

\begin{lstlisting}[language=c++, float=*htbp, caption={A truncated ROOT interactive session displaying the contents of an object of the \emph{GeFiCa::X} class.}, label={l:x}]
  root [] GeFiCa::X x
  (GeFiCa::X &) Name: x Title: 1D Cartesian coordinate
  root [] x.Dump()
  ==> Dumping object at: 0x00007f76e5d80150, name=x, class=GeFiCa::X
  Src                      ->7f76e5d802a8      -(net impurity concentration)x|Qe|/epsilon
  N1                       101                 number of points along the 1st coordinate
  N2                       0                   number of points along the 2nd coordinate
  N3                       0                   number of points along the 3rd coordinate
  MaxIterations            5000                maximal iterations of SOR to be performed
  RelaxationFactor         1.95                within (0,2), used to speed up convergence
  Precision                1e-07               difference between two consecutive SOR iterations
  ...
\end{lstlisting}

Macros are organized in sub-folders in \emph{GeFiCa/examples/} to demonstrate the usage of GeFiCa classes. The \emph{planar/}, \emph{trueCoaxial/}, \emph{hemispherical/}, \emph{pointContact/}, and \emph{segmented/} folders are used to show how to configure specific types of HPGe detectors and then calculate the fields in them. The \emph{analytic/} and the \emph{fenics/} folders contains macros that are independent of the \emph{GeFiCa} libraries. The macros in the former demonstrate how to calculate and visualize the field distribution in simple HPGe detectors using ROOT. The latter shows Python codes to calculate and visualize the field distribution in a simplified point-contact geometry using FEniCS~\cite{fenics}. All field distributions shown in this work are generated using these macros. A user can learn the topics by at first running these macros to reproduce plots in this work, and then modifying them to meet his/her own needs.

\section{Code Verification}
A common way to verify the saneness of a complex theory in physics is to consider extreme conditions, under which the theory can be simplified and compared to predictions based on common sense. Take the field in a point-contact detector as an example, there are two extreme cases where the field in certain part of the detector can be regarded as the same as that in a planar or a true-coaxial detector. This makes it possible to compare the numeric calculation of a point-contact detector field directly with analytic solutions.

\subsection{Comparison with Analytic Solutions}
\begin{figure}[htbp]
  \includegraphics[width=\linewidth]{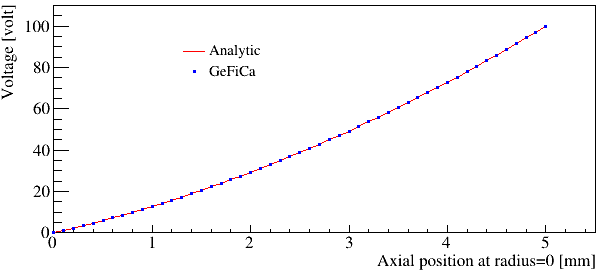}
  \includegraphics[width=\linewidth]{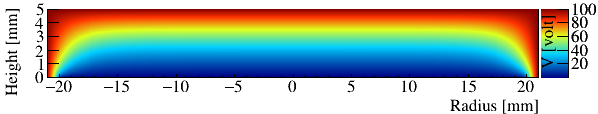}
  \caption{Top: Comparison of the electric potential calculated numerically in a pancake-like ``point-contact'' detector with the analytic solution of a planar detector that has the same thickness (or height) and impurity concentration. Bottom: The electric potential distribution calculated numerically in the pancake-like detector, the ``point-contact'' of which is artificially enlarged to cover almost the entire bottom end surface.}
  \label{f:pancake}
\end{figure}

In the first extreme case we consider a point-contact detector that takes a pancake-like shape, that is, its thickness is much smaller than its diameter. Furthermore, its ``point-contact'' covers almost the entire bottom end surface. The electric potential in such a detector is shown in the bottom plot in Fig.~\ref{f:pancake}. At the radial center of the detector, the field is essentially the same as that in a planar detector that has the same thickness and impurity concentration. In the top plot in Fig.~\ref{f:pancake} the analytic solution of such a planar detector is overlaid on top of the numerical result of the pancake-like ``point-contact'' detector along the axial positions at radius = 0.

\begin{figure}[htbp]\centering
  \includegraphics[height=0.36\textheight]{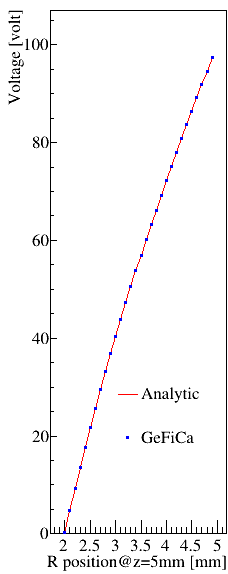}
  \includegraphics[height=0.36\textheight]{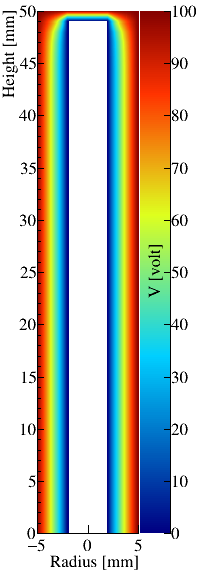}
  \caption{Left: Comparison of the electric potential calculated numerically in a thin-stick-like ``point-contact'' detector with the analytic solution of a true-coaxial detector that has the same radius and impurity concentration. Right: The electric potential distribution calculated numerically in the thin-stick-like detector, the ``point-contact'' of which is artificially prolonged along almost the entire height of the crystal.}
  \label{f:match}
\end{figure}

In the second case let us consider a point-contact detector that looks like a thin stick, that is, its diameter is much smaller than its height. Furthermore, let's make its ``point-contact'' as deep into the crystal as possible. The potential distribution in such a detector calculated numerically is shown in the right plot in Fig.~\ref{f:match}. Far away from the top end of the detector, the field is essentially the same as that in a true-coaxial detector that has the same radius and impurity concentration. In the left plot in Fig.~\ref{f:match} the analytic solution of such a true-coaxial detector is overlaid on top of the numerical result of the thin-stick-like ``point-contact'' detector along the radius at an axial position 5~mm above the bottom surface.

\subsection{Comparison with Fieldgen}
Even though the perfect matches between the analytic solutions and the numerical results in both cases are convincing evidences of the correctness of the numerical calculation implemented in GeFiCa, it is worth noting that a constant impurity concentration throughout the entire crystal is assumed to make the analytic solutions possible. In case of an arbitrary impurity distribution, no simple analytic solution is available, the numerical calculation in GeFiCa is compared to that of fieldgen~\cite{siggen, ringberg}, a thoroughly examined and widely accepted package in the field, given identical point-contact detector configurations.

\begin{figure}[htbp]\centering
  \includegraphics[width=0.7\linewidth]{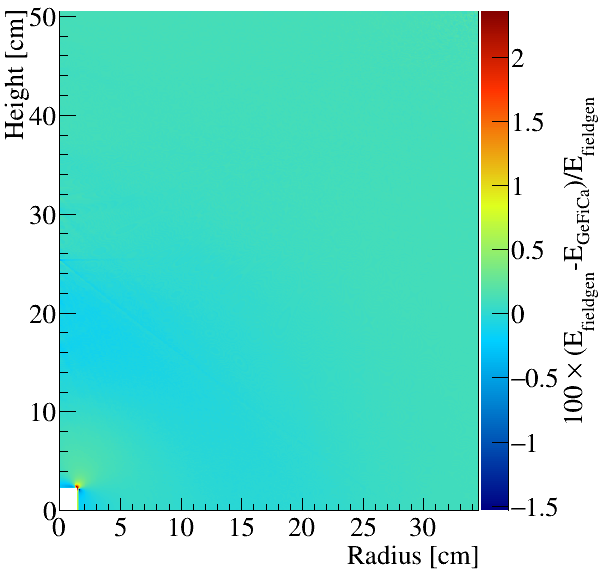}
  \caption{Relative difference between the electric potential distributions calculated using fieldgen and GeFiCa for an identical point-contact detector configuration.}
  \label{f:dpc}
\end{figure}

The biggest difference between GeFiCa and fieldgen in the aspect of numerical calculation is probably the setup of grid points. In case of fieldgen, the grid points along the radial direction, $r$, of a detector start from $r=0$ and end at $r=$ the radius of the detector. In case of GeFiCa, the grid points are in the range of [$-$radius, +radius] and there is no grid point at $r=0$ to avoid setting artificial boundary conditions at $r=0$. Due to this difference, there is no one-to-one correspondence between a grid point in GeFiCa and a grid point in fieldgen. In order to make a point-to-point comparison, linear interpolation is used to get the total electric field strength at a fieldgen point from two nearby GeFiCa points, the interpolated value is then compared to the fieldgen value at the same point. Their relative difference in percentage is shown as colored contour in Fig.~\ref{f:dpc}.

The largest difference is about 8.5\% at the top right corner of the point-contact. This point is removed from Fig.~\ref{f:dpc} so that subtle differences between fieldgen and GeFiCa are more visible in the figure. The second largest difference is about 2.5\% at an adjacent point, shown as the red spot in Fig.~\ref{f:dpc}. The difference quickly falls below 0.1\% only a few points away from the corner, which translates to about one mini-meter in length given the 0.1~mm distance between grid points. Such difference is most probably due to different treatments in fieldgen and GeFiCa on grid points near boundaries.

Fortunately, the difference is of little importance in practice since there is no such sharp corner inside any detector in reality. Predictions of GeFiCa and fieldgen in the bulk of the detector are essentially identical.

\section{Performance}
\subsection{Relaxation Factor}
\label{s:rf}
\begin{figure}[htbp]\centering
    \includegraphics[width=\linewidth]{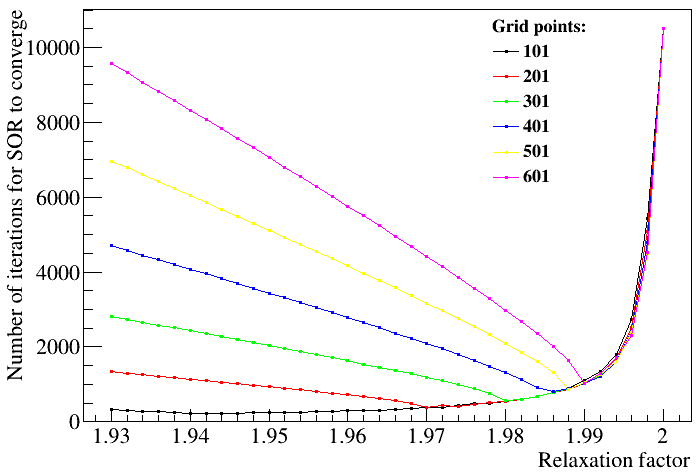}
    \caption{Number of SOR iterations versus relaxation factor.}
    \label{f:rf}
\end{figure}

As described in Sec.~\ref{s:sor}, the number of iterations needed for a successive relaxation process to converge can be reduced by introducing a relaxation factor in between [1,2). Fig.~\ref{f:rf} shows the number of iterations for a succesive over-relaxation (SOR) process to converge as a function of the relaxation factor. Each data point in the figure represents the result from a numerical calculation of the field in an ideal planar detector. Data points that are connected by lines in between are from calculations sharing the same number of grid points.

A common trend shared by all the lines is that there is a point where the number of iterations is minimized. That is where GeFiCa reaches its best performance. As the number of grid points increase from 101 to 601, the relaxation factor corresponding to the minimal iteration numbers increases from around 1.94 to 1.99. The default value of the relaxation factor is set to 1.95 in GeFiCa. A user can change it using the following line of code if desired.
\begin{lstlisting}[language=c++]
  grid.RelaxationFactor=1.99;
\end{lstlisting}

The gain in performance by selecting an appropriate relaxation factor becomes more prominent when the number of grid points becomes larger. Take the up most curve in Fig.~\ref{f:rf} as an example, which corresponds to calculations done with the finest grid, when the relaxation factor changes by only 0.06 from 1.93 to 1.99, the number of iterations reduces from about 10,000 to less than 2,000. While the lowest curve in Fig.~\ref{f:rf} is almost flat around 1.94, that is, the relaxation factor cannot help much to gain speed for calculations with a very coarse grid. This is not a problem since those calculations are fast already.

After every 100 iterations, GeFiCa prints the overall difference of potentials at all grid points between current and previous iterations. When the difference is smaller than a target precision ($1\times10^{-7}$~V by default), the SOR is regarded as converged, the calculation stops there, and the CPU time used for the calculatioin is printed out on screen as shown in the terminal output below:
\begin{lstlisting}[basicstyle=\small]
root [0] .x calculateFields.cc
Processing calculateFields.cc...
Info in <GeFiCa::RhoZ::SuccessiveOverRelax>:
Start...
   0 steps, precision: 1.0e+00 (target: 1e-07)
 100 steps, precision: 4.8e-03 (target: 1e-07)
 200 steps, precision: 2.7e-03 (target: 1e-07)
   .
   .
   .
2000 steps, precision: 1.0e-07 (target: 1e-07)
2004 steps, precision: 1.0e-07 (target: 1e-07)
Info in <GeFiCa::RhoZ::SuccessiveOverRelax>:
CPU time: 23.2 s
\end{lstlisting}
This terminal output is associated with the calculation used to generate Fig.~\ref{f:dpc}. The overall number of grid points is 349,140. The CPU time used for the calculation is about 23 second in a Linux server with an Intel Xeon Gold 5118 CPU at 1~GHz. The relaxation factor chosen for this calculation is 1.994.


\subsection{Output File Size}
\label{s:fs}
The output of fieldgen used to generate Fig.~\ref{f:dpc} is saved as a simple ASCII file that is 8.1 Mega bytes in size. The detector configuration is saved as a short header of the file. The rest of the file are six columns of values of the grid point positions (radial and axial), the voltage, the overall electric field strength, and its radial and axial components.

As described in Sec.~\ref{s:dsio}, the detector and grid objects in GeFiCa can be directly saved in a standard ROOT file. Its contents can be printed and visualized in a ROOT interactive session as demonstrated in the code snippets in Sec.~\ref{s:dsio} and \ref{s:ms}. In addition to the information saved in a fieldgen output, a GeFiCa output also contains the intervals between grid points, flags indicating whether a point is depleted or not, etc. It also contains about twice more grid points than fieldgen. In total, the amount of information saved in GeFiCa is about 4 times more than that saved in the fieldgen output. The size of the GeFiCa output ROOT file used to generate Fig.~\ref{f:dpc} is 9.2 Mega bytes, only slightly larger than that of the fieldgen ouput file, thanks to the \emph{gzip} algrithm used to compress a ROOT file mentioned in Sec.~\ref{s:dsio}.

\section{Extendability and Limitation}
\begin{figure}[htbp]
  \centering
  \includegraphics[width=0.8\linewidth]{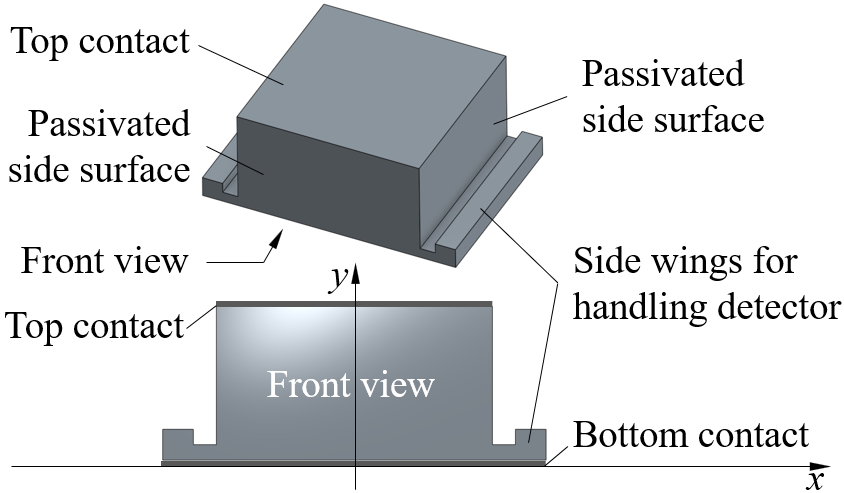}
  \caption{Configuration of a realistic planar detector that has two side wings for the handing of the detector. The top and bottom surfaces are electrical contacts, the side surfaces are passivated.}
  \label{f:hat}
\end{figure}

Let's take a realistic planar detector configuration shown in Fig.~\ref{f:hat} as an example to demonstrate the procedure of extending GeFiCa for a new type of detector.

The top and bottom surfaces of the detector are covered with a thin layer of aluminium to form the electric contacts. All the side surfaces are covered with a thin layer of amorphous germanium for passivation purpose. The two side wings can be used for handling the detector without touching its sensitive surfaces~\cite{amman18}. Since they are thin compared to the overall thickness of the detector, the electric field distribtuion inside the detector can hence be approximated by that in an ideal 1D planar detector. However, if our intention is to study the influence of the thickness of the wings on the electric field, we need at least a 2D grid in Cartesian coordinates to perform the numerical calculation, which can be achieved with the following steps.

At first, a class called \emph{XY} that represents the dimension and coordinates needs to be created. It inherits all member variables in its base class \emph{Grid} that define the grid. Since the numerical expression of Poisson's Equation (Eq.~\ref{e:pxyz}) depends on dimensions and coordinates used for the calculation, a protected virtual function, \emph{void OverRelaxAt(size\_t idx)}, in \emph{Grid} needs to be overwritten in \emph{XY}, which takes care of the updating of the field value at each grid point indexed by \emph{idx}.

Secondly, a class called \emph{TopHat} that describes the geometry of the detector needs to be created. It inherits the member variables that hold voltages values of all electrodes from its base class \emph{Detector}. It also inherits the impurity distribution from the class, \emph{crystal}. A public member function \emph{void Draw()} in \emph{Detector} needs to be overwritten in {TopHat} to visualize the geometry setup.

At last the public virtual function in \emph{Grid} called \emph{void SetupWith(Detector\&)} needs to be overwritten in \emph{XY}, which takes the boundary conditions and impurity distribution from \emph{TopHat} to construct and initialize the grid for the calculation.

For completeness, a folder called \emph{TopHat} is recommended to be created under \emph{GeFiCa/examples}, which contains ROOT or Python scripts demonstrating the usage of \emph{XY} and \emph{TopHat}.

Given its extendability, there is no limitation on GeFiCa from the functionality point of view. From the education point of view, however, there is currently no function in GeFiCa demonstrating the adaptive grid configuration that automatically updates distances between grid points over iterations based on the strength of local electric field. Note that there is no fundamental limitation from GeFiCa inhibiting doing so, since there are separated member variables in the \emph{grid} class to hold distances from a grid point to its neighbors in all directions. Practically, GeFiCa is already fast and precise enough with fixed step length for common HPGe configurations. This function can be added in if necessary.

\section{Summary}
The new educational program, GeFiCa, has been created to demonstrate analytic and numeric methods to calculate static electric fields and potentials in HPGe detectors. It is freely available from \url{http://physino.xyz/gefica} and can be installed in three major operating systems, Linux, MacOS and Windows, as a CERN ROOT~\cite{root} library extension. Powered by ROOT, GeFiCa allows its users to explore in detail the calculation procedure by executing C++ or Python code snippets in ROOT interactive sessions or Jupyter notebooks without compilation. Example code snippets are shipped together with the library to demonstrate calculations for common detector configurations, and to visualize the resulting field distributions in graphs or color contours. In addition to field calculations, GeFiCa offers functionalities to calculate the HPGe detector depletion voltage, undepleted region, capacitance, etc., that are not available from general-purpose field calculation programs, such as Maxwell3D and FEniCS. Compared to open projects that are also specialized in HPGe field calculation, such as fieldgen and SSD, etc., GeFiCa offers a ROOT-based C++ solution that is equally accurate and efficient, and shipped with a large amount of documentations and examples that are not readily available in others.

This article was written to provide an entry level review of methods and tools available at the moment, with the hope that its readers feel comfortable to make an educated choice of simulation tools best suited for the task at their hands.

\begin{acknowledgements}
  The authors thank David Radford at the Oak Ridge National Laboratory for his patient instruction in various aspects of the field calculation, Oliver Schulz at the Max-Planck-Institut f\"ur Physik for his introduction of the Julia language and the SSD package, Christopher Haufe and Anna Reine at the University of North Carolina at Chapel Hill for their instruction on how to use FEniCS to calculate fields in a point-contact detector. This work is supported by NSF award OIA-1738695 and OISE-1743790, and the Office of Research at the University of South Dakota. Computations supporting this project were performed on High Performance Computing systems at the University of South Dakota, funded by NSF award OAC-1626516.
\end{acknowledgements}

\appendix

\section{Poisson's Equation in Curvilinear Coordinates}
\label{a:pesc}
\begin{figure}[htbp]
  \includegraphics[width=0.48\linewidth]{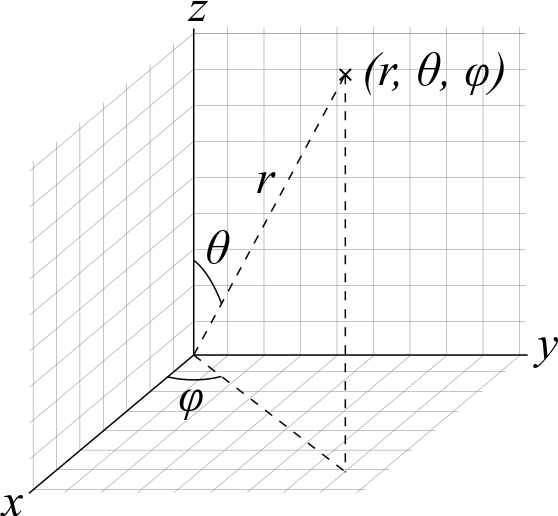}
  \includegraphics[width=0.5\linewidth]{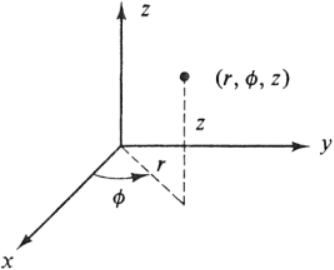}
  \caption{Definitions of spherical coordinates (left) and cylindrical ones (right).}
  \label{f:sphp}
\end{figure}

The Poisson's equation in spherical coordinates reads,
\begin{align}
  \begin{split}
    &\frac{1}{r^2}\frac{\partial}{\partial r}\left(r^2\frac{\partial V}{\partial r}\right)
    +\frac{1}{r^2\sin\theta}\frac{\partial V}{\partial\theta} \left(\sin\theta\frac{\partial V}{\partial\theta} \right)\\
    &+\frac{1}{r^2\sin^2\theta}\frac{\partial V^2}{\partial^2\phi}
    =-\frac{\rho(r,\theta,\phi)}{\epsilon},
  \end{split}
\end{align}
where $r, \theta \in [0, \pi], \phi \in [0,2\pi)$ are the radial distance from the origin, the polar angle and the azimuth angle, as defined in the left plot of Fig.~\ref{f:sphp}.

The Poisson's equation in cylindrical coordinates reads,
\begin{equation}
  \frac{1}{r}\frac{\partial}{\partial r}\left(r\frac{\partial V}{\partial r}\right)
  +\frac{1}{r^2}\frac{\partial V^2}{\partial^2\phi}
  +\frac{\partial V^2}{\partial^2z}
  =-\frac{\rho(r,\phi,z)}{\epsilon},
\end{equation}
where $r, \phi \in [0,2\pi), z$ are the radial distance from the origin, the azimuth angle, and the height, as defined in the right plot of Fig.~\ref{f:sphp}. A more commonly used symbol of the radial distance in the cylindrical coordinate system is $\rho$. However, $r$ is used here instead of $\rho$ to avoid being confused with the space charge density, which is denoted as $\rho$ as well.

\section{Iteration relations}
\label{a:3dc}
In 3D Cartesian coordinates, the potential at a grid point after the $i$-th successive relaxation iteration, $V_{i+1}$, can be expressed as

\begin{eqnarray*}
  V_{i+1}&=&\left\lbrace\frac{\rho}{2\epsilon}\right.\\
    &+&\left[\frac{ V_i(x+dx_+)}{dx_+}+\frac{ V_i(x+dx_-)}{dx_-}\right]
	\frac{1}{dx_++dx_-}\\
    &+&\left[\frac{ V_i(y+dy_+)}{dy_+}+\frac{ V_i(y+dy_-)}{dy_-}\right]
	\frac{1}{dy_++dy_-}\\
    &+&\left[\frac{ V_i(z+dz_+)}{dz_+}+\frac{ V_i(z+dz_-)}{dz_-}\right]
	\frac{1}{dz_++dz_-}
    \left.\vphantom{\frac12}\right\rbrace\\
    &/&\left[\left(\frac{1}{dx_+}+\frac{1}{dx_-}\right)\left(\frac{1}{dx_++dx_-}\right)\right.\\
    &+&\left(\frac{1}{dy_+}+\frac{1}{dy_-}\right)\left(\frac{1}{dy_++dy_-}\right)\\
    &+&\left(\frac{1}{dz_+}+\frac{1}{dz_-}\right)\left(\frac{1}{dz_++dz_-}\right)
    \left.\vphantom{\frac12}\right],
\end{eqnarray*}
where $x, y, z$ are coordinates, $dx_+, dy_+, dz_+$ are distances to the next grid points, $dx_-, dy_-, dz_-$ distances to the previous.

In a 1D cylindrical coordinate, the potential at a grid point after the $i$-th successive relaxation iteration, $V_{i+1}$, can be expressed as

\begin{eqnarray*}
  V_{i+1} &=& \frac{\rho}{\epsilon} \frac{\dd r_+ \dd r_-}{2}\\
  &+&\frac{V_i(r+\dd r_+)-V_i(r-\dd r_-)}{2r}/\left(\frac{1}{\dd r_+}+\frac{1}{\dd r_-}\right)\\
  &+&\left[\frac{V_i(r+\dd r_+)}{\dd r_+} + \frac{V_i(r-\dd r_-)}{\dd r_-}\right]
  /\left(\frac{1}{\dd r_+}+\frac{1}{\dd r_-}\right),
\end{eqnarray*}
where $r$ is the coordinate. $dr_+$ is the distance to the next grid point, $dr_-$ the distance to the previous.

In 3D cylindrical coordinate, the potential at a grid point after the $i$-th successive relaxation iteration, $V_{i+1}$, can be expressed as

\begin{eqnarray*}
  V_{i+1}&=&\left\lbrace\frac{\rho}{2\epsilon}+\frac{1}{r}\frac{V_i(r+\dd r_+)-V_i(r+\dd r_-)}{\dd r_++\dd r_-}\right.\\
    &+&\left[\frac{ V_i(r+\dd r_+)}{\dd r_+}+\frac{ V_i(r+\dd r_-)}{\dd r_-}\right]
\frac{1}{\dd r_++\dd r_-}\\
    &+&\left[\frac{ V_i(\theta+d\theta_+)}{d\theta_+}+\frac{ V_i(\theta+d\theta_-)}{d\theta_-}\right]
\frac{1}{d\theta_++d\theta_-}\frac{1}{r^2}\\
    &+&\left[\frac{ V_i(z+dz_+)}{dz_+}+\frac{ V_i(z+dz_-)}{dz_-}\right]
\frac{1}{dz_++dz_-}
    \left.\vphantom{\frac12}\right\rbrace\\
    &/&\left[\left(\frac{1}{\dd r_+}+\frac{1}{\dd r_-}\right)\left(\frac{1}{\dd r_++\dd r_-}\right)\right.\\
    &+&\left(\frac{1}{d\theta_+}+\frac{1}{d\theta_-}\right)\left(\frac{1}{d\theta_++d\theta_-}\frac{1}{r^2}\right)\\
    &+&\left(\frac{1}{dz_+}+\frac{1}{dz_-}\right)\left(\frac{1}{dz_++dz_-}\right)
    \left.\vphantom{\frac12}\right],
\end{eqnarray*}
where $r, \theta, z$ are coordinates, $\dd r_+, \dd\theta_+, \dd z_+,$ are step lengths to the next grid points, $\dd r_-, \dd\theta_-, \dd z_-$ step lengths to the previous.

In a 1D spherical coordinate, the potential at a grid point after the $i$-th successive relaxation iteration, $V_{i+1}$, can be expressed as
\begin{eqnarray*}
  V_{i+1} &=& \frac{\rho}{\epsilon} \frac{\dd r_+ \dd r_-}{2}\\
  &+&\frac{V_i(r+\dd r_+)-V_i(r-\dd r_-)}{r}/\left(\frac{1}{\dd r_+}+\frac{1}{\dd r_-}\right)\\
  &+&\left[\frac{V_i(r+\dd r_+)}{\dd r_+} + \frac{V_i(r-\dd r_-)}{\dd r_-}\right]
  /\left(\frac{1}{\dd r_+}+\frac{1}{\dd r_-}\right),
\end{eqnarray*}
where $r$ is the coordinate, $\dd r_+$ is the distance to the next grid point, $\dd r_-$ the distance to the previous.

In 3D Spherical Coordinate, the potential at a grid point after the $i$-th successive relaxation iteration, $V_{i+1}$, can be expressed as
\begin{eqnarray*}
  V_{i+1}&=&\left\lbrace\frac{\rho}{2\epsilon}\right.\\
     &+&\frac{2}{r}\frac{V_i(r+\dd r_+)-V_i(r+\dd r_-)}{\dd r_++\dd r_-}\\
     &+&\frac{1}{r^2\tan\theta}\frac{V_i(\theta+d\theta_+)-V_i(\theta+d\theta_-)}{d\theta_++d\theta_-}\\
     &+&\left[\frac{V_i(r+\dd r_+)}{\dd r_+}+\frac{V_i(r+\dd r_-)}{\dd r_-}\right] \frac{1}{\dd r_++\dd r_-}\\
     &+&\left[\frac{V_i(\theta+d\theta_+)}{d\theta_+} + \frac{V_i(\theta+d\theta_-)}{d\theta_-}\right] \frac{1}{d\theta_++d\theta_-}\frac{1}{r^2}\\
     &+&\left[\frac{V_i(\phi+d\phi_+)}{d\phi_+} + \frac{ V_i(\phi+d\phi_-)}{d\phi_-}\right] \frac{1}{dz_++dz_-}\frac{1}{r^2\sin^2\theta} \left.\vphantom{\frac12}\right\rbrace\\
     &/&\left[\left(\frac{1}{\dd r_+}+\frac{1}{\dd r_-}\right)\left(\frac{1}{\dd r_++\dd r_-}\right)\right.\\
     &+&\left(\frac{1}{d\theta_+}+\frac{1}{d\theta_-}\right)\left(\frac{1}{d\theta_++d\theta_-}\frac{1}{r^2}\right)\\
     &+&\left(\frac{1}{d\phi_+}+\frac{1}{d\phi_-}\right)\left(\frac{1}{d\phi_++d\phi_-}\right)\frac{1}{r^2\sin^2\theta} \left.\vphantom{\frac12}\right],
\end{eqnarray*}
where $r, \theta, \phi$ are coordinates, $\dd r_+, \dd\theta_+, \dd\phi_+,$ are the step lengths to the next grid points, $\dd r_-, \dd\theta_-, \dd\phi_-$ to the previous.

\bibliography{ref}
\bibliographystyle{spphys}
\end{document}